\documentstyle[10pt,epsf,epsfig,hangcaption,xspace,amssymb,amsfonts,amsmath,amsthm,cite,
dp_delphititle,lineno]{dp_delphi}
\makeindex
\def\DpPaperGroup{EP}
\def\DpPaperRef{2003-058}
\def\DpDate{2 September 2003}
\def\DpAuthors{DELPHI Collaboration}
\def\DpSubmit{(Accepted by Eur. Phys. J. C)}
\def\DpTitle{{Study of Tau-pair Production in 
Photon-Photon Collisions at LEP and 
Limits on the Anomalous Electromagnetic 
Moments of the Tau Lepton}}
\def\DpComment{  }
\def\DpEMail{  }

\newcommand{\etal}{{\it et al.,\/}\ }
\newcommand{\signal}{${\mathrm e}^+{\mathrm e}^- \rightarrow {\mathrm e}^+{\mathrm e}^-\tau^+\tau^-$}
\newcommand{\degrees}{^\circ}
\newcommand{\dedx}{d{\it E}/d{\it x} }
\newfont{\scsl}{ecsc1200} 
\newcommand{\Zfitter}{{\scsl{%
\raisebox{-0.4ex}{z\kern-0.05em{}f}\kern-0.1em{}I\kern-0.15em%
\raisebox{0.8ex}{T}\kern-0.25em{}T\kern-0.25em%
\raisebox{-0.8ex}{E\kern-0.05em{}r}}}}
\begin{document}
\makeatletter
\newcount\@tempcntc
\def\@citex[#1]#2{\if@filesw\immediate\write\@auxout{\string\citation{#2}}\fi
  \@tempcnta\z@\@tempcntb\m@ne\def\@citea{}\@cite{\@for\@citeb:=#2\do
    {\@ifundefined
       {b@\@citeb}{\@citeo\@tempcntb\m@ne\@citea\def\@citea{,}{\bf ?}\@warning
       {Citation `\@citeb' on page \thepage \space undefined}}%
    {\setbox\z@\hbox{\global\@tempcntc0\csname b@\@citeb\endcsname\relax}%
     \ifnum\@tempcntc=\z@ \@citeo\@tempcntb\m@ne
       \@citea\def\@citea{,}\hbox{\csname b@\@citeb\endcsname}%
     \else
      \advance\@tempcntb\@ne
      \ifnum\@tempcntb=\@tempcntc
      \else\advance\@tempcntb\m@ne\@citeo
      \@tempcnta\@tempcntc\@tempcntb\@tempcntc\fi\fi}}\@citeo}{#1}}
\def\@citeo{\ifnum\@tempcnta>\@tempcntb\else\@citea\def\@citea{,}%
  \ifnum\@tempcnta=\@tempcntb\the\@tempcnta\else
   {\advance\@tempcnta\@ne\ifnum\@tempcnta=\@tempcntb \else \def\@citea{--}\fi
    \advance\@tempcnta\m@ne\the\@tempcnta\@citea\the\@tempcntb}\fi\fi}
 
\makeatother
\begin{titlepage}
\pagenumbering{roman}
\CERNpreprint{\DpPaperGroup}{\DpPaperRef} 
\date{{\small\DpDate}} 
\title{\DpTitle} 
\address{\DpAuthors} 
\begin{shortabs} 
\noindent
\noindent
Tau-pair production in the process \signal\  was studied using data collected 
by the DELPHI experiment at LEP2 during the years 1997 -- 2000. 
The corresponding integrated
luminosity is 650 pb$^{-1}$. The values of the cross-section obtained are 
found to be in agreement with QED predictions. Limits on the anomalous 
magnetic and electric dipole moments of the tau lepton are deduced.
\end{shortabs}
\vfill
\begin{center}
\DpSubmit \ \\ 
\DpComment \ \\
\DpEMail \ \\
\end{center}
\vfill
\clearpage
\headsep 10.0pt
\addtolength{\textheight}{10mm}
\addtolength{\footskip}{-5mm}
\begingroup
%
\newcommand{\DpName}[2]{\hbox{#1$^{\ref{#2}}$},\hfill}
\newcommand{\DpNameTwo}[3]{\hbox{#1$^{\ref{#2},\ref{#3}}$},\hfill}
\newcommand{\DpNameThree}[4]{\hbox{#1$^{\ref{#2},\ref{#3},\ref{#4}}$},\hfill}
\newskip\Bigfill \Bigfill = 0pt plus 1000fill
\newcommand{\DpNameLast}[2]{\hbox{#1$^{\ref{#2}}$}\hspace{\Bigfill}}
%
\footnotesize
\noindent
\DpName{J.Abdallah}{LPNHE}
\DpName{P.Abreu}{LIP}
\DpName{W.Adam}{VIENNA}
\DpName{P.Adzic}{DEMOKRITOS}
\DpName{T.Albrecht}{KARLSRUHE}
\DpName{T.Alderweireld}{AIM}
\DpName{R.Alemany-Fernandez}{CERN}
\DpName{T.Allmendinger}{KARLSRUHE}
\DpName{P.P.Allport}{LIVERPOOL}
\DpName{U.Amaldi}{MILANO2}
% ...
\normalsize
\endgroup
\titlefoot{Department of Physics and Astronomy, Iowa State
     University, Ames IA 50011-3160, USA
    \label{AMES}}
\titlefoot{Physics Department, Universiteit Antwerpen,
     Universiteitsplein 1, B-2610 Antwerpen, Belgium \\
     \indent~~and IIHE, ULB-VUB,
     Pleinlaan 2, B-1050 Brussels, Belgium \\
     \indent~~and Facult\'e des Sciences,
     Univ. de l'Etat Mons, Av. Maistriau 19, B-7000 Mons, Belgium
    \label{AIM}}
\titlefoot{Physics Laboratory, University of Athens, Solonos Str.
     104, GR-10680 Athens, Greece
    \label{ATHENS}}
\titlefoot{Department of Physics, University of Bergen,
     All\'egaten 55, NO-5007 Bergen, Norway
    \label{BERGEN}}
\titlefoot{Dipartimento di Fisica, Universit\`a di Bologna and INFN,
     Via Irnerio 46, IT-40126 Bologna, Italy
    \label{BOLOGNA}}
\titlefoot{Centro Brasileiro de Pesquisas F\'{\i}sicas, rua Xavier Sigaud 150,
     BR-22290 Rio de Janeiro, Brazil \\
     \indent~~and Depto. de F\'{\i}sica, Pont. Univ. Cat\'olica,
     C.P. 38071 BR-22453 Rio de Janeiro, Brazil \\
     \indent~~and Inst. de F\'{\i}sica, Univ. Estadual do Rio de Janeiro,
     rua S\~{a}o Francisco Xavier 524, Rio de Janeiro, Brazil
    \label{BRASIL}}
\titlefoot{Coll\`ege de France, Lab. de Physique Corpusculaire, IN2P3-CNRS,
     FR-75231 Paris Cedex 05, France
    \label{CDF}}
\titlefoot{CERN, CH-1211 Geneva 23, Switzerland
    \label{CERN}}
\titlefoot{Institut de Recherches Subatomiques, IN2P3 - CNRS/ULP - BP20,
     FR-67037 Strasbourg Cedex, France
    \label{CRN}}
\titlefoot{Now at DESY-Zeuthen, Platanenallee 6, D-15735 Zeuthen, Germany
    \label{DESY}}
\titlefoot{Institute of Nuclear Physics, N.C.S.R. Demokritos,
     P.O. Box 60228, GR-15310 Athens, Greece
    \label{DEMOKRITOS}}
\titlefoot{FZU, Inst. of Phys. of the C.A.S. High Energy Physics Division,
     Na Slovance 2, CZ-180 40, Praha 8, Czech Republic
    \label{FZU}}
\titlefoot{Dipartimento di Fisica, Universit\`a di Genova and INFN,
     Via Dodecaneso 33, IT-16146 Genova, Italy
    \label{GENOVA}}
\titlefoot{Institut des Sciences Nucl\'eaires, IN2P3-CNRS, Universit\'e
     de Grenoble 1, FR-38026 Grenoble Cedex, France
    \label{GRENOBLE}}
\titlefoot{Helsinki Institute of Physics, P.O. Box 64,
     FIN-00014 University of Helsinki, Finland
    \label{HELSINKI}}
\titlefoot{Joint Institute for Nuclear Research, Dubna, Head Post
     Office, P.O. Box 79, RU-101 000 Moscow, Russian Federation
    \label{JINR}}
\titlefoot{Institut f\"ur Experimentelle Kernphysik,
     Universit\"at Karlsruhe, Postfach 6980, DE-76128 Karlsruhe,
     Germany
    \label{KARLSRUHE}}
\titlefoot{Institute of Nuclear Physics,Ul. Kawiory 26a,
     PL-30055 Krakow, Poland
    \label{KRAKOW1}}
\titlefoot{Faculty of Physics and Nuclear Techniques, University of Mining
     and Metallurgy, PL-30055 Krakow, Poland
    \label{KRAKOW2}}
\titlefoot{Universit\'e de Paris-Sud, Lab. de l'Acc\'el\'erateur
     Lin\'eaire, IN2P3-CNRS, B\^{a}t. 200, FR-91405 Orsay Cedex, France
    \label{LAL}}
\titlefoot{School of Physics and Chemistry, University of Lancaster,
     Lancaster LA1 4YB, UK
    \label{LANCASTER}}
\titlefoot{LIP, IST, FCUL - Av. Elias Garcia, 14-$1^{o}$,
     PT-1000 Lisboa Codex, Portugal
    \label{LIP}}
\titlefoot{Department of Physics, University of Liverpool, P.O.
     Box 147, Liverpool L69 3BX, UK
    \label{LIVERPOOL}}
\titlefoot{Dept. of Physics and Astronomy, Kelvin Building,
     University of Glasgow, Glasgow G12 8QQ
    \label{GLASGOW}}
\titlefoot{LPNHE, IN2P3-CNRS, Univ.~Paris VI et VII, Tour 33 (RdC),
     4 place Jussieu, FR-75252 Paris Cedex 05, France
    \label{LPNHE}}
\titlefoot{Department of Physics, University of Lund,
     S\"olvegatan 14, SE-223 63 Lund, Sweden
    \label{LUND}}
\titlefoot{Universit\'e Claude Bernard de Lyon, IPNL, IN2P3-CNRS,
     FR-69622 Villeurbanne Cedex, France
    \label{LYON}}
\titlefoot{Dipartimento di Fisica, Universit\`a di Milano and INFN-MILANO,
     Via Celoria 16, IT-20133 Milan, Italy
    \label{MILANO}}
\titlefoot{Dipartimento di Fisica, Univ. di Milano-Bicocca and
     INFN-MILANO, Piazza della Scienza 2, IT-20126 Milan, Italy
    \label{MILANO2}}
\titlefoot{IPNP of MFF, Charles Univ., Areal MFF,
     V Holesovickach 2, CZ-180 00, Praha 8, Czech Republic
    \label{NC}}
\titlefoot{NIKHEF, Postbus 41882, NL-1009 DB
     Amsterdam, The Netherlands
    \label{NIKHEF}}
\titlefoot{National Technical University, Physics Department,
     Zografou Campus, GR-15773 Athens, Greece
    \label{NTU-ATHENS}}
\titlefoot{Physics Department, University of Oslo, Blindern,
     NO-0316 Oslo, Norway
    \label{OSLO}}
\titlefoot{Dpto. Fisica, Univ. Oviedo, Avda. Calvo Sotelo
     s/n, ES-33007 Oviedo, Spain
    \label{OVIEDO}}
\titlefoot{Department of Physics, University of Oxford,
     Keble Road, Oxford OX1 3RH, UK
    \label{OXFORD}}
\titlefoot{Dipartimento di Fisica, Universit\`a di Padova and
     INFN, Via Marzolo 8, IT-35131 Padua, Italy
    \label{PADOVA}}
\titlefoot{Rutherford Appleton Laboratory, Chilton, Didcot
     OX11 OQX, UK
    \label{RAL}}
\titlefoot{Dipartimento di Fisica, Universit\`a di Roma II and
     INFN, Tor Vergata, IT-00173 Rome, Italy
    \label{ROMA2}}
\titlefoot{Dipartimento di Fisica, Universit\`a di Roma III and
     INFN, Via della Vasca Navale 84, IT-00146 Rome, Italy
    \label{ROMA3}}
\titlefoot{DAPNIA/Service de Physique des Particules,
     CEA-Saclay, FR-91191 Gif-sur-Yvette Cedex, France
    \label{SACLAY}}
\titlefoot{Instituto de Fisica de Cantabria (CSIC-UC), Avda.
     los Castros s/n, ES-39006 Santander, Spain
    \label{SANTANDER}}
\titlefoot{Inst. for High Energy Physics, Serpukov
     P.O. Box 35, Protvino, (Moscow Region), Russian Federation
    \label{SERPUKHOV}}
\titlefoot{J. Stefan Institute, Jamova 39, SI-1000 Ljubljana, Slovenia
     and Laboratory for Astroparticle Physics,\\
     \indent~~Nova Gorica Polytechnic, Kostanjeviska 16a, SI-5000 Nova Gorica, Slovenia, \\
     \indent~~and Department of Physics, University of Ljubljana,
     SI-1000 Ljubljana, Slovenia
    \label{SLOVENIJA}}
\titlefoot{Fysikum, Stockholm University,
     Box 6730, SE-113 85 Stockholm, Sweden
    \label{STOCKHOLM}}
\titlefoot{Dipartimento di Fisica Sperimentale, Universit\`a di
     Torino and INFN, Via P. Giuria 1, IT-10125 Turin, Italy
    \label{TORINO}}
\titlefoot{INFN,Sezione di Torino, and Dipartimento di Fisica Teorica,
     Universit\`a di Torino, Via P. Giuria 1,\\
     \indent~~IT-10125 Turin, Italy
    \label{TORINOTH}}
\titlefoot{Dipartimento di Fisica, Universit\`a di Trieste and
     INFN, Via A. Valerio 2, IT-34127 Trieste, Italy \\
     \indent~~and Istituto di Fisica, Universit\`a di Udine,
     IT-33100 Udine, Italy
    \label{TU}}
\titlefoot{Univ. Federal do Rio de Janeiro, C.P. 68528
     Cidade Univ., Ilha do Fund\~ao
     BR-21945-970 Rio de Janeiro, Brazil
    \label{UFRJ}}
\titlefoot{Department of Radiation Sciences, University of
     Uppsala, P.O. Box 535, SE-751 21 Uppsala, Sweden
    \label{UPPSALA}}
\titlefoot{IFIC, Valencia-CSIC, and D.F.A.M.N., U. de Valencia,
     Avda. Dr. Moliner 50, ES-46100 Burjassot (Valencia), Spain
    \label{VALENCIA}}
\titlefoot{Institut f\"ur Hochenergiephysik, \"Osterr. Akad.
     d. Wissensch., Nikolsdorfergasse 18, AT-1050 Vienna, Austria
    \label{VIENNA}}
\titlefoot{Inst. Nuclear Studies and University of Warsaw, Ul.
     Hoza 69, PL-00681 Warsaw, Poland
    \label{WARSZAWA}}
\titlefoot{Fachbereich Physik, University of Wuppertal, Postfach
     100 127, DE-42097 Wuppertal, Germany
    \label{WUPPERTAL}}
\addtolength{\textheight}{-10mm}
\addtolength{\footskip}{5mm}
\clearpage
\headsep 30.0pt
\end{titlepage}
%
\pagenumbering{arabic} 
\setcounter{footnote}{0} %
\large
\section{Introduction}
This paper presents a study of tau pair production in photon-photon 
collisions using the data collected by the DELPHI detector
at LEP in the period from 1997 to 2000 (LEP2) 
at collision energy $\sqrt{s}$ between 183 and 208 GeV.
The total integrated luminosity used in the analysis is 650 pb$^{-1}$.
At LEP this process was first 
observed by the OPAL collaboration \cite{opal_ggtt} and 
subsequently studied by the L3 collaboration \cite{l3_ggtt}.

\begin{figure}[h]
  \begin{center}
    \mbox{\epsfig{file=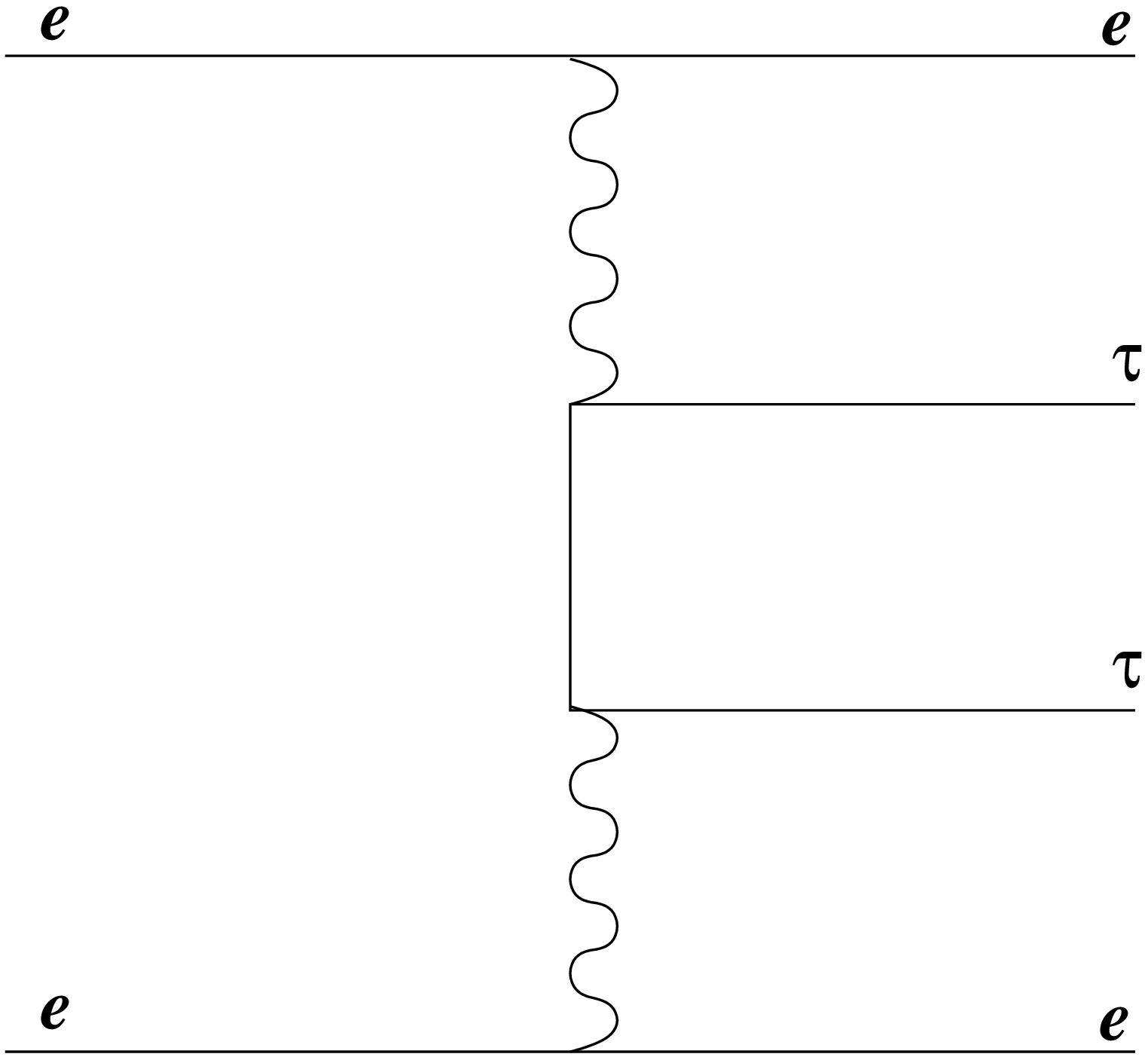,width=0.32\textwidth}}
    \caption{The dominant diagram for the reaction \signal\ }
    \label{diagram}
  \end{center}
\end{figure}

The final state ${\mathrm e^+e^-}\tau^+\tau^-$ can be produced via a set of Feynman
diagrams. In this paper we present the cross-section measurement
for the contribution of the so-called multiperipheral graph 
(Fig. \ref{diagram}) which corresponds to collisions of two virtual 
photons. The same final states produced via other diagrams
(less then 1\% of the cross-section)
are considered as a background.

The study of the reaction \signal\  explores two fundamental problems.
First of all it provides a deep test of QED at the level of the fourth order in 
$\alpha$. Furthermore,  
the $\gamma\tau\tau$ vertex is sensitive to the anomalous electromagnetic 
couplings of the tau lepton. Since the multiperipheral 
\signal\  process diagram 
contains two such vertices, the anomalous magnetic and electric dipole
moments can be extracted by comparing the measured cross-section with
QED expectations.

The rest of the paper gives a detailed description of the 
\signal\  cross-section measurement, including tau-pair selection, 
background estimation, selection and trigger efficiency calculation
and systematic error estimation. 
In the last part of the paper the measured cross-sections
are used to derive limits on the 
anomalous electromagnetic moments of the tau lepton.

\section{Monte Carlo simulation}
The signal process was simulated using the Berends, Daverveldt and Kleiss 
generator RADCOR (BDKRC) \cite{bdkrc}, which calculates the cross-section for
the multiperipheral diagram with  radiative corrections on the electron and
positron lines. The following signal definition was used: 
the invariant mass of tau pairs
had to be less than 40 GeV/c$^2$; both beam particles had to be scattered 
by less than 10 degrees; and at least one of them had to be scattered
by less than 2 degrees. 
With these restrictions the accepted
cross-section was 1.44$\pm$0.04\% lower than the total cross-section
predicted by BDKRC (which is about 450 pb at LEP2 energies)
for the unrestricted phase space. 
The $\tau$ decay was simulated by the TAUOLA 
package \cite{tauola},
which includes photon radiation from the decay products.   
The BDKRC generator was also used to estimate the background coming from the 
process ${ \mathrm e^+e^- \rightarrow e^+e^-}\mu^+\mu^-$. 

To simulate the ${\mathrm e^+e^- \rightarrow e^+e^-e^+e^-}$ background, 
the Berends, Daverveldt and Kleiss generator DIAG36 (BDK) \cite{bdk} was used. 
Hadron production in two photon collisions was simulated by 
PYTHIA 6.1 \cite{pythia}. 
Non-multiperipheral four-fermion processes (such as WW, ZZ, Zee and others) were simulated by
WPHACT \cite{wphact}. 

The generated events were passed through the full simulation program of the DELPHI 
detector and were reconstructed with the same program  as for
the real data \cite{delphi}.

\section{Event selection}
In most events produced by two-photon collisions
both beam particles scatter at small angles and remain 
undetectable inside the beam pipe. Therefore only the decay products of the tau 
leptons can be seen in the detector. To suppress background, only one-prong
decay channels with one tau decaying into an electron and the other into a
non-electron (hadron or muon) were considered. 
The analysis was based entirely on the measured tracks
of charged products of tau decays; the neutral particles
from tau decays were ignored in this analysis.


To select runs with good performance of the sub-detectors 
\cite{delphi}--\cite{delphi1}, only runs with the Time Projection Chamber (TPC),
the Forward Chambers (FCA, FCB) and one of the additional barrel 
tracking detectors (ID or VD) fully operational were retained.
Table~\ref{lumi} presents the luminosities used in the analysis,
luminosity-weighted centre-of-mass energies and energy ranges.

\begin{table}[h]
  \begin{center}
    \begin{tabular}{|l|c|c|c|c|} 
      \hline 
                       & 1997 & 1998 & 1999           & 2000           \\
      \hline
 Luminosity, pb$^{-1}$ & 52.3 & 152.6& 224.2          & 217.5          \\
       $<E_{cm}>$, GeV &182.7 & 188.7& 197.6          & 206.3          \\
Energy range, GeV      &182.7 & 188.7& 195.5 -- 201.5 & 204.5 -- 208.0 \\
       \hline
     \end{tabular}
   \caption{The integrated luminosities, mean collision energies and collision energy ranges.}
   \label{lumi}
 \end{center}
\end{table}

The event selection procedure was divided into two steps. The first step 
(preselection) selected a sample of two-photon events with two good tracks
which were not  back-to-back in the plane perpendicular to the beam axis.
A track was 
considered as good if the 
momentum derived from its curvature was greater than 100 MeV/c,
momentum error better than 100\%, 
polar angle $\theta$ between $20\degrees$ and $160\degrees$,
and impact parameter with respect to the interaction point below 
10 cm along the $z$-axis  
\footnote{The DELPHI coordinate system has the $z$-axis aligned along the electron beam direction,
  the $x$-axis pointing toward the centre of LEP and the $y$-axis vertical. 
  $r$ is the
  radius in the $(x,y)$ plane. The polar angle $\theta$ is measured with 
  respect to the $z$-axis and the azimuthal angle $\phi$ is about $z$.}
and 5 cm in the $r-\phi$ plane. 

The following cuts were applied in this first selection step:
\begin{itemize}
\item{There had to be exactly two good tracks from particles with opposite charges, 
      at least one of them having momentum greater than 300 MeV/c.}
\item{To suppress background from fermion pair production, 
      the total energy of two charged particles
      had to be less than 30 GeV.}
\item{To enrich the sample with \signal\  events, 
      the acoplanarity of two tracks
\footnote{Acoplanarity is defined as $180^{\circ}-|\phi_2-\phi_1|$.} 
had to be greater than $0.5\degrees$ and their resultant transverse momentum 
greater than 500 MeV/c.
}
\item{ To select events with a high trigger efficiency, the transverse energy,
   defined by $$ E_t = E_1\sin{\theta_1}+E_2\sin{\theta_2},$$
   where $E_1$ and $E_2$ are the energies of the two charged particles
   and $\theta_1$ and $\theta_2$ are their polar angles,
   had to be greater than 2 GeV. } 
\item{In the year 2000, the operation of one of the twelve TPC sectors 
   was unstable and the \dedx measurement vital for this analysis was poor,
   so events with at least one track in or near (closer than 10$^\circ$ in $\phi$)
   to this TPC sector in 2000 were rejected. }
\item{ Finally, to ensure the transverse momentum balance of $\gamma\gamma$ system, 
 single and double tagged events were rejected by requiring that no energy
deposition in the forward electromagnetic calorimeters (STIC or FEMC) exceeded 60\% of the beam
energy.}
\end{itemize}

\begin{figure}[h]
  \begin{center}
    \begin{tabular}{c c}
      \mbox{\epsfig{file=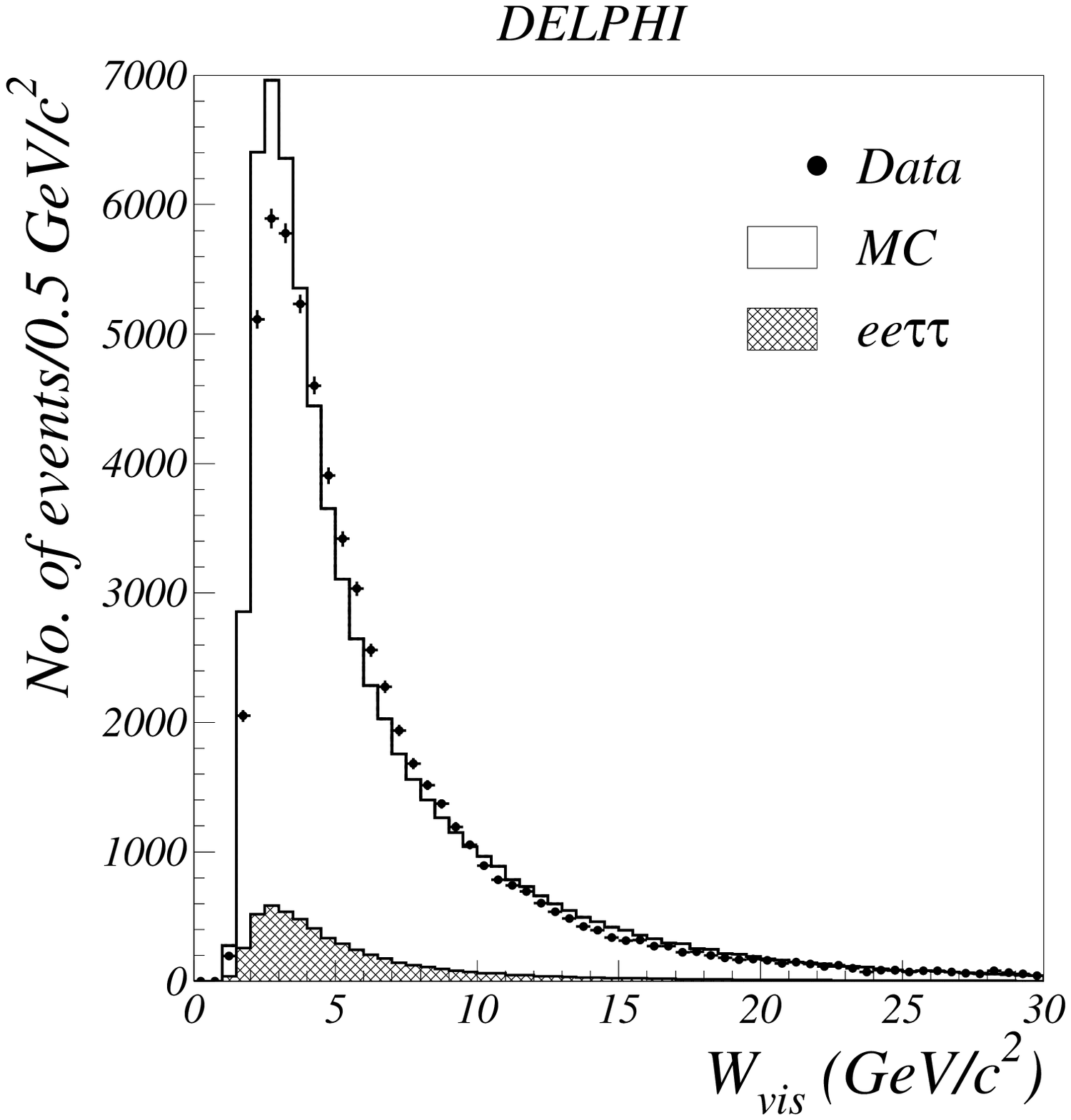,width=0.48\textwidth}} & 
      \mbox{\epsfig{file=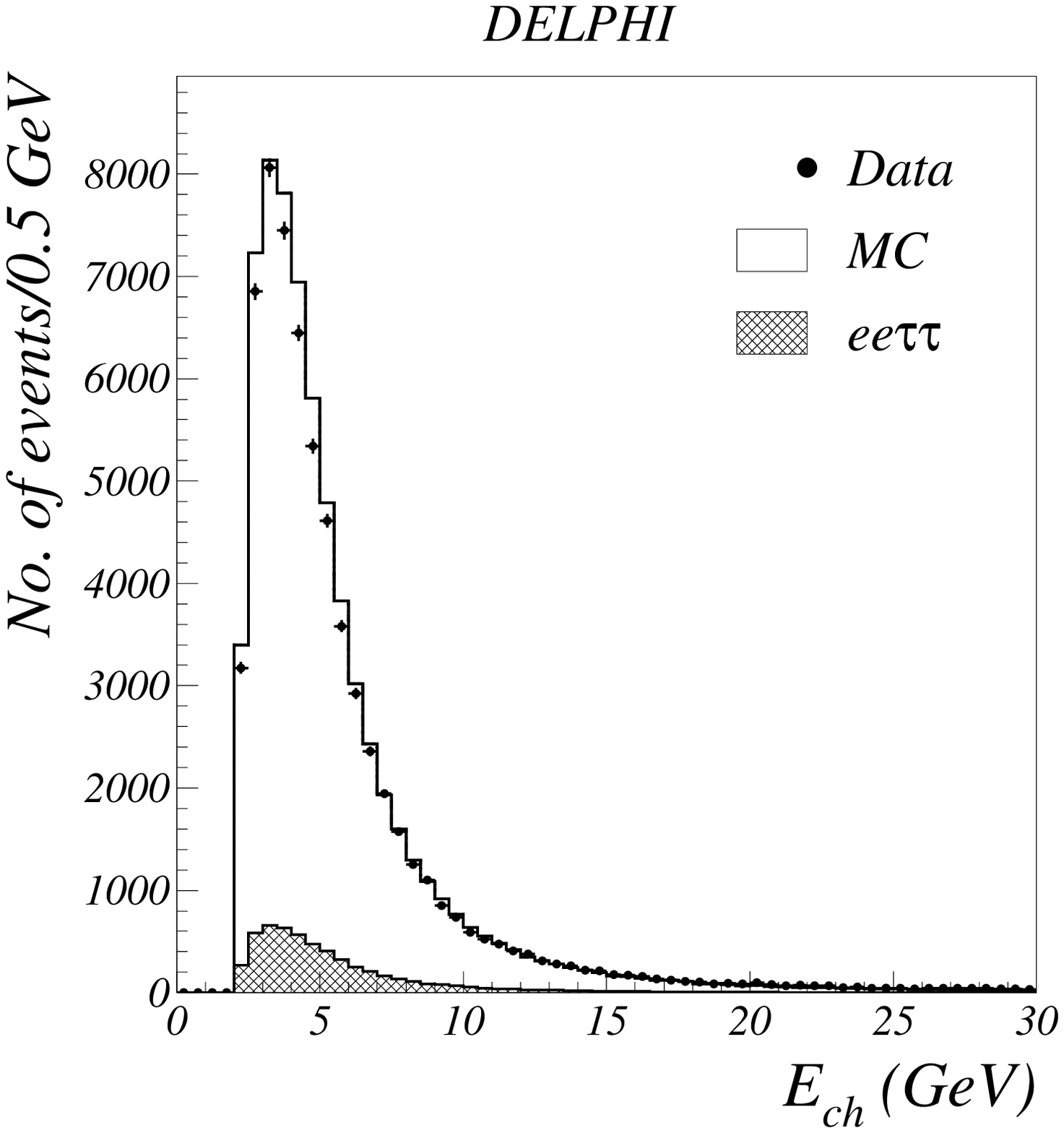,width=0.48\textwidth}}\\
      \mbox{\epsfig{file=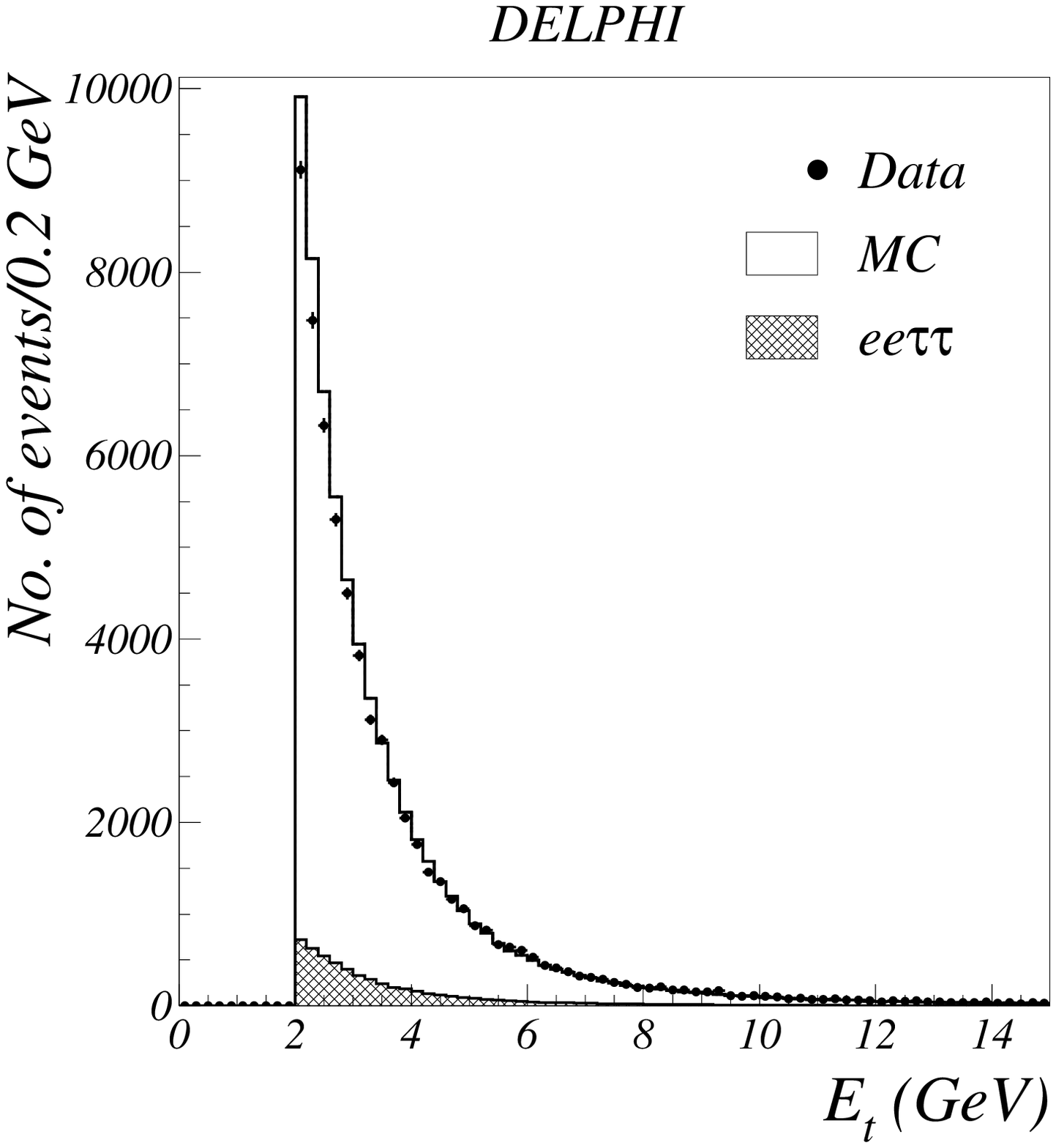,width=0.48\textwidth}} & 
      \mbox{\epsfig{file=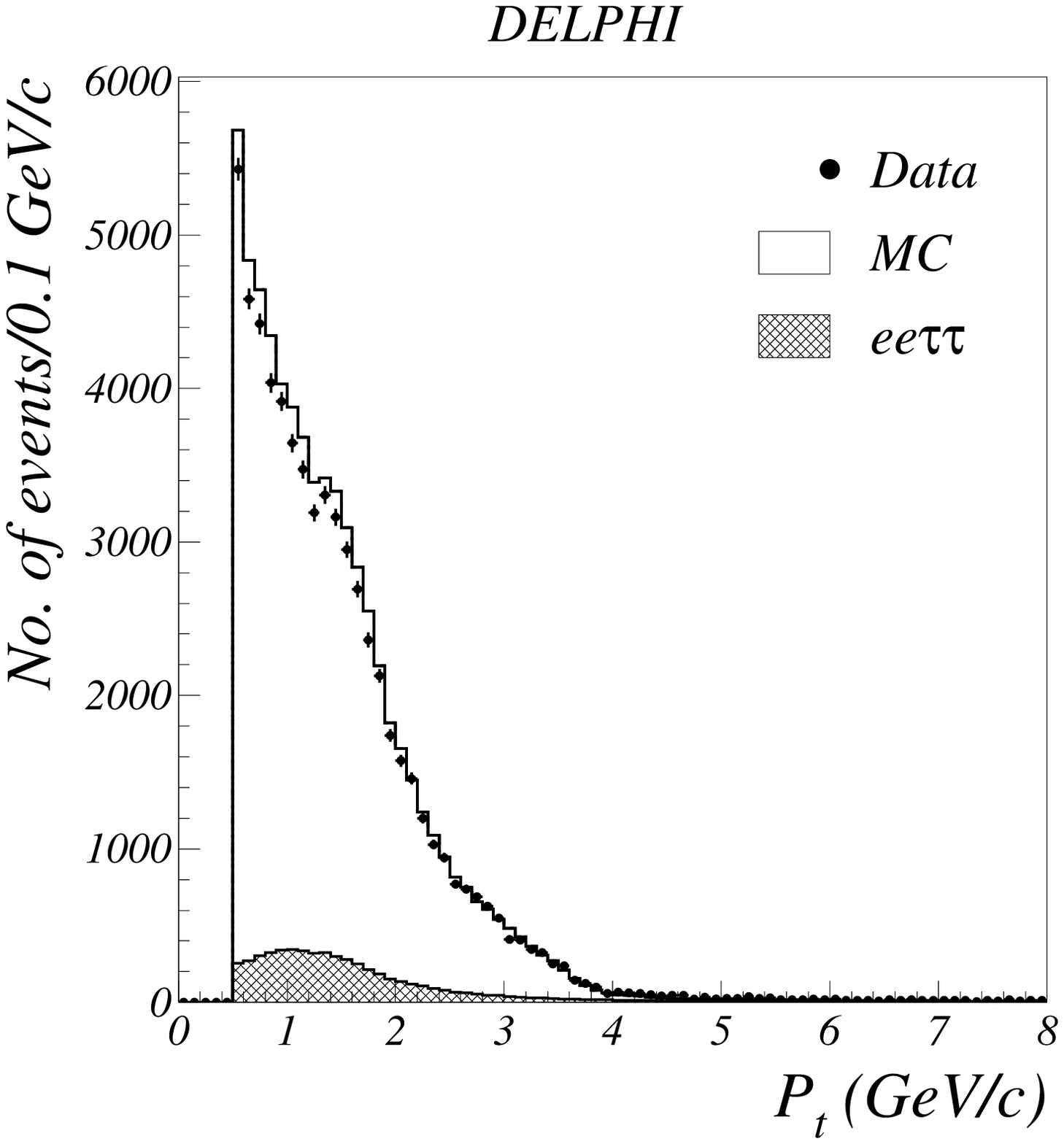,width=0.48\textwidth}}
    \end{tabular}
  \end{center}
  \caption{The distributions of invariant mass, total energy, transverse energy
    and transverse momentum of the pair of charged particles.
    Preselection cuts are applied. The points are 1999 data, the open histogram 
    is the simulation of background processes and the shaded histogram is the
    simulation of \signal\  events. The simulation is not corrected for the trigger efficiency.}
  \label{preselection}
\end{figure}

The last cut suppressed the events with highly virtual photons.
About 90\% of events passing this cut had the momentum transfer $-q^2$ 
less then 1 GeV$^2/c^2$.
After applying the cuts described above, the predicted event composition in the
preselected sample  was as follows (1999 data):
\begin{center}
  \begin{tabular}{ll}
    ${ \mathrm e^+e^- \rightarrow e^+e^-e^+e^-}$     & 41 \% \\
    ${ \mathrm e^+e^- \rightarrow e^+e^-}\mu^+\mu^-$ & 47 \% \\
    \signal\                                &  8 \% \\
    ${ \mathrm e^+e^- \rightarrow e^+e^-q\bar{q}}$   &  3 \% \\
    ${ \mathrm e^+e^-} \rightarrow \tau^+\tau^-$     &  1 \% \\
  \end{tabular}
\end{center}
The fraction of other events was less than 1\%.
The efficiency of the preselection for \signal\ events was of the order of 5\%,
the largest suppression of the signal coming from
the requirement of exactly two good tracks 
seen in the detector
(about a factor of 4) and from the cut on $E_t$ (about factor of 2).
Figure \ref{preselection} shows the comparison between data and simulation  
of the distributions of invariant mass, total energy, 
total transverse energy and total transverse momentum 
of the pair of charged particles.
The \signal\  events are shown by the shaded histogram.
The Monte Carlo is normalised to the luminosity of the real data.
The data deficit is mainly due to the trigger inefficiency
which is corrected at the later stages of analysis.

In the final step of the selection, the event was retained if one 
of the charged particles was identified as an electron and the other as a non-electron.
This step was based on the TPC measurement of the \dedx 
pulls for the muon, electron, kaon and proton 
hypotheses. The \dedx pull for a specific particle hypothesis is defined
as the ratio
\begin{equation}
\Pi_X = \frac{({\mathrm d}E/{\mathrm d}x)_{meas}-({\mathrm d}E/{\mathrm d}x)_{exp}}
    {\sigma_{{\mathrm d}E/{\mathrm d}x}},
\end{equation}
where $({\mathrm d}E/{\mathrm d}x)_{exp}$ is the value expected for the particle 
$X$ with given momentum
and $\sigma_{{\mathrm d}E/{\mathrm d}x}$ is the error of the measured energy loss 
$({\mathrm d}E/{\mathrm d}x)_{meas}$.
To check the \dedx calibration, test samples of 
${ \mathrm e^+e^- \rightarrow e^+e^-}\mu^+\mu^-$ and 
${ \mathrm e^+e^- \rightarrow e^+e^-e^+e^-}$ events 
were picked out from the preselected sample.
A small angular dependence of the \dedx measurements was found as well as some disagreement 
between data and simulation.
Corrections which were functions of azimuthal and polar angle were applied to
the measured \dedx values.
Residual disagreement was removed by scaling and smearing the specific energy loss
measurement in the simulated events. 
Independent calibrations of real and simulated data were performed for each year of data taking 
analysed.
The efficiency to measure \dedx is discussed later in the paper.

\begin{figure}[h]
\begin{center}
\begin{tabular}{c c}
 \mbox{\epsfig{file=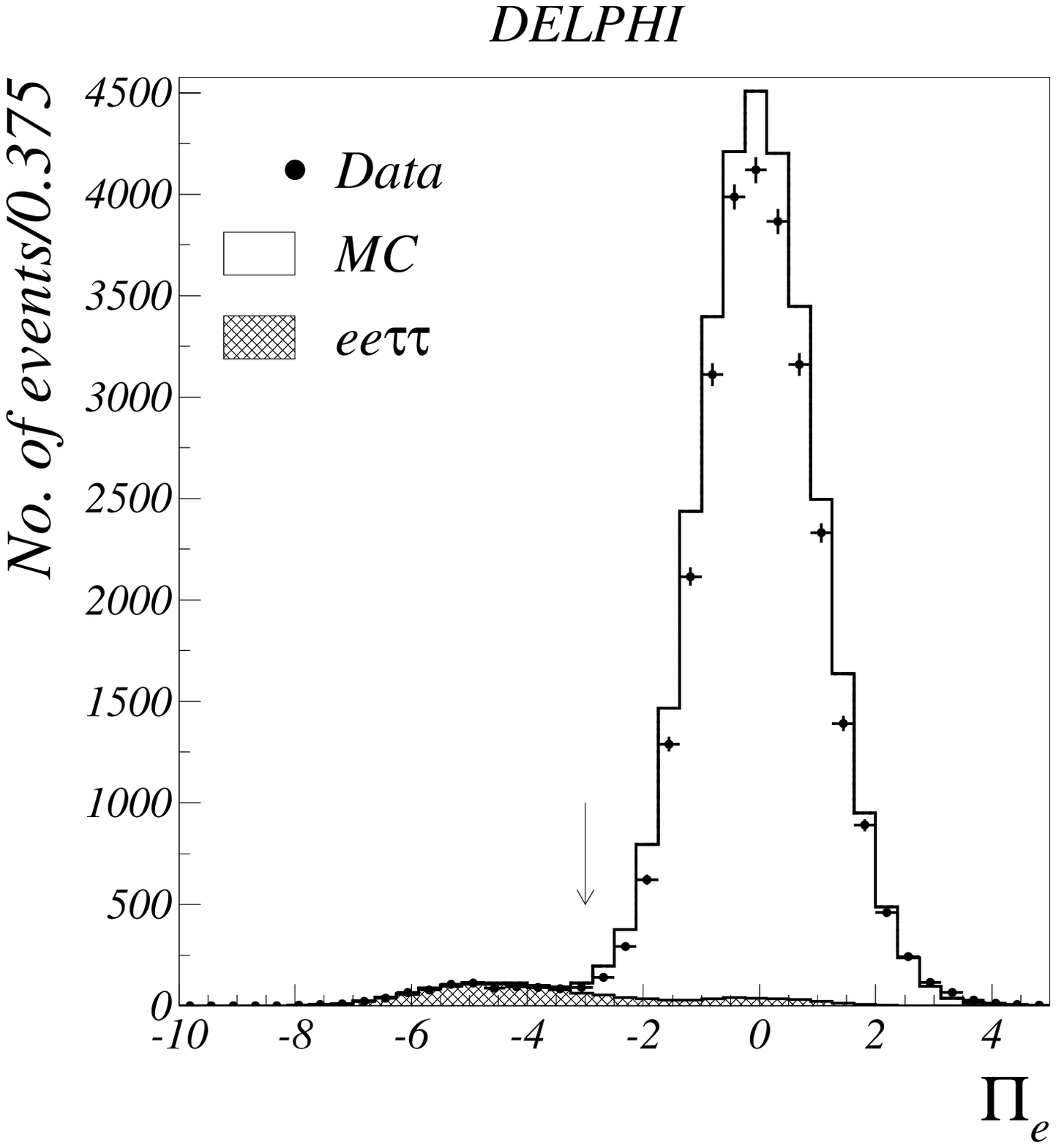,width=0.48\textwidth}} 
 \mbox{\epsfig{file=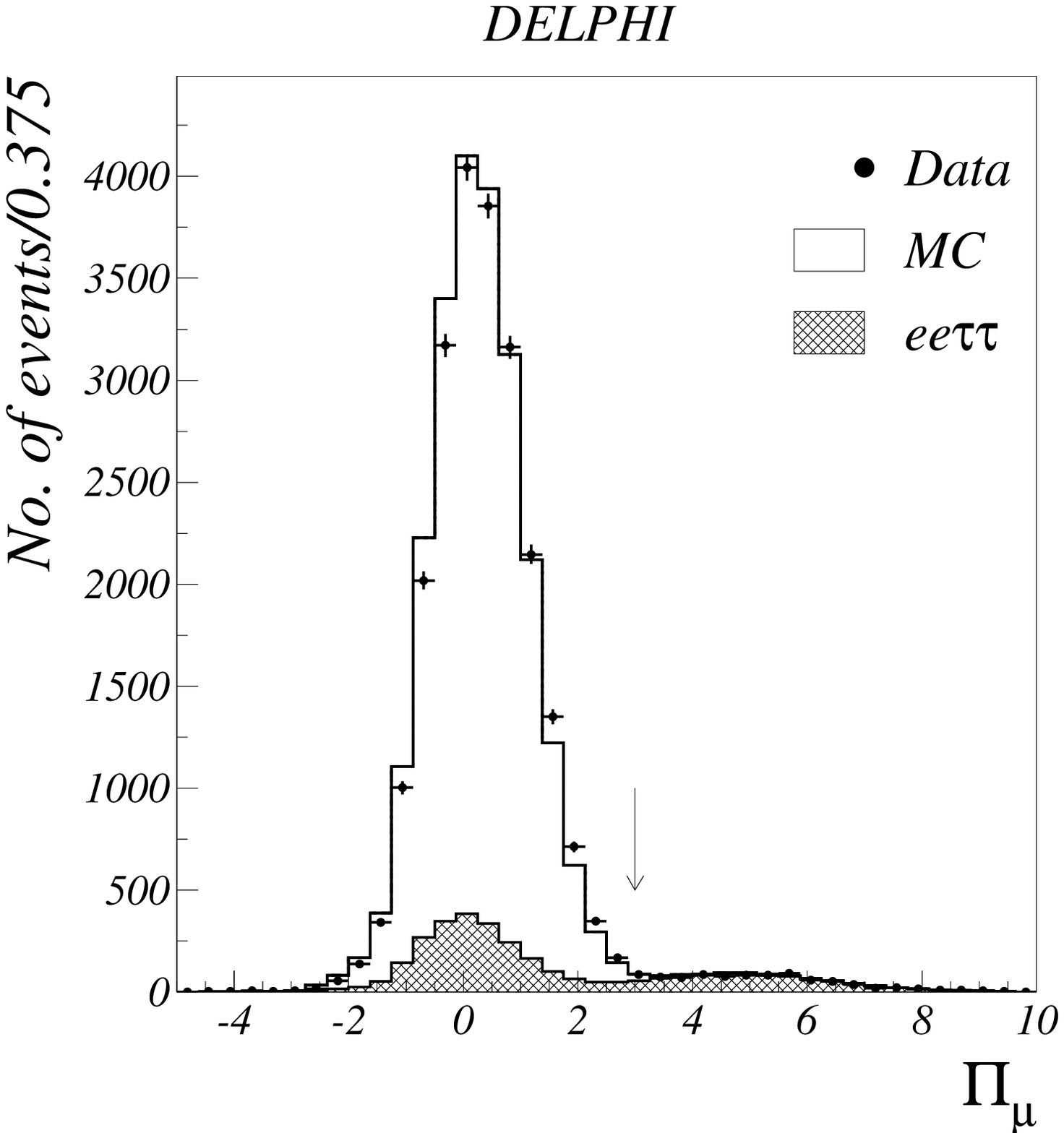,width=0.48\textwidth}} 
\end{tabular}

\caption{ Distribution of the \dedx pull for the electron hypothesis (left) and 
  for the muon hypothesis (right) with all selection cuts applied except the
  cut on the variable shown. Points are 1999 data, the open histogram is the 
  background, and the shaded histogram shows the \signal\ signal events.
  The simulation is not corrected for the trigger efficiency.}
\label{pulls}
\end{center}
\end{figure}

With corrected \dedx information, a track was identified as an electron if $\Pi_{\mu}>3$
and as a non-electron if $\Pi_{\mathrm e}<-3$. 
Figure \ref{pulls} illustrates the particle identification cuts. The distributions of the
pulls for the electron and muon hypotheses are shown for the 1999 real data and simulation. 
Each distribution is shown after applying all selection cuts except the cut on the
variable shown. The shaded histograms show the signal. 

\begin{figure}[h]
\begin{center}
\begin{tabular}{c c}
 \mbox{\epsfig{file=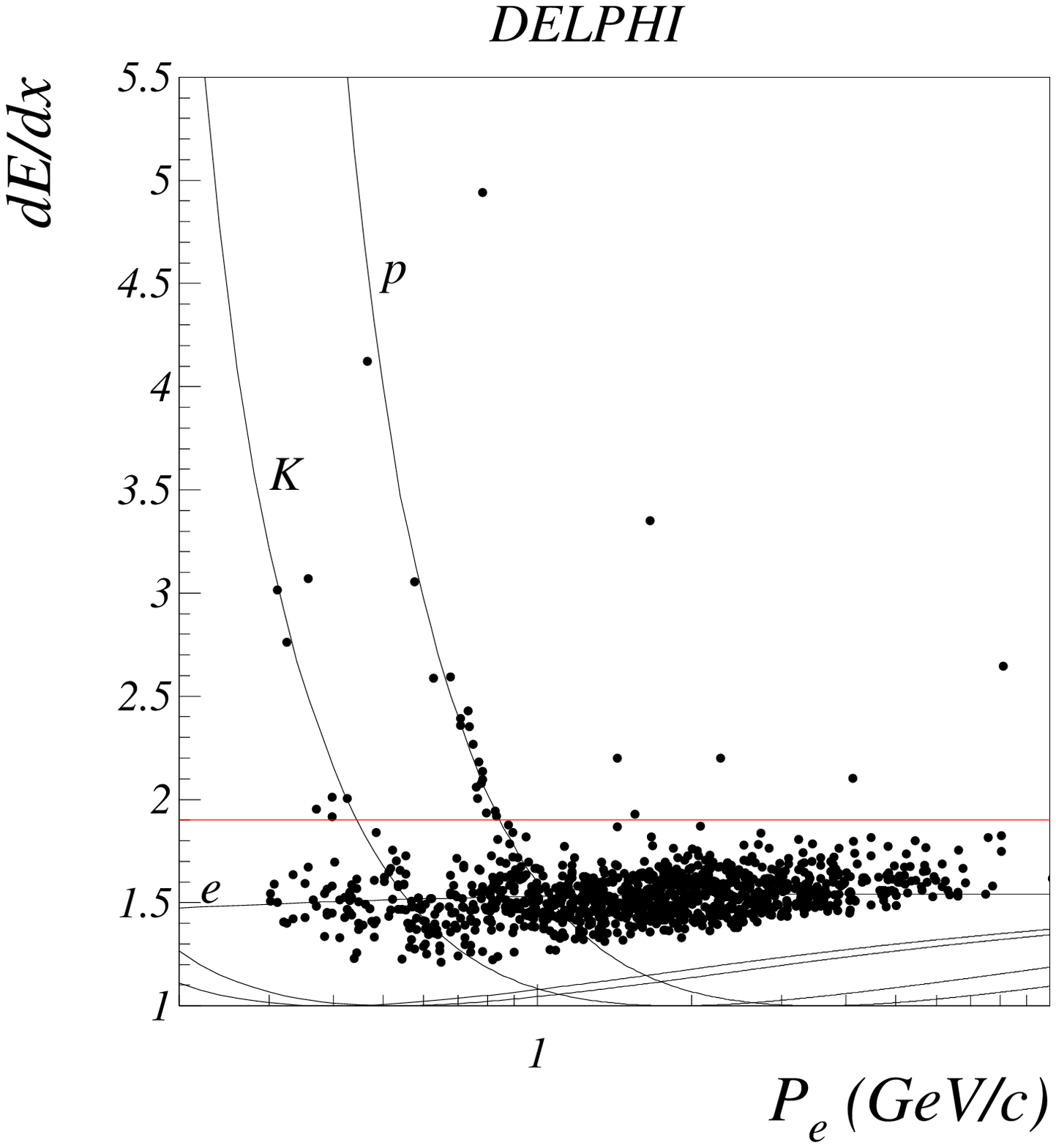,width=0.48\textwidth}}
 \mbox{\epsfig{file=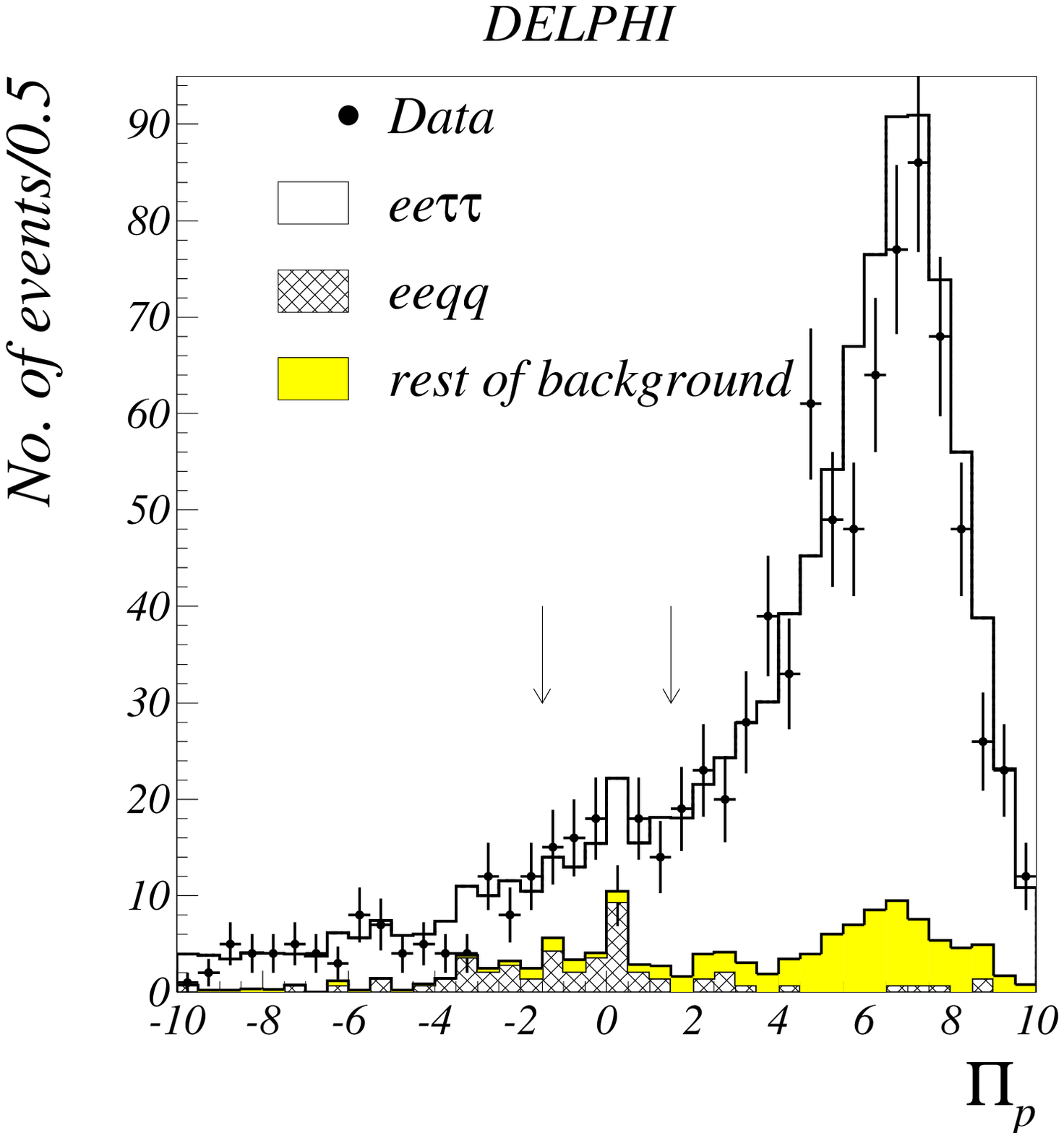,width=0.48\textwidth}}
\end{tabular} 
\caption{Left: specific energy loss as a function of particle momentum 
  for electron candidates after preselection cuts and the cuts on the 
  electron and muon pulls. 
  The horizontal line shows the first 
  cut against kaons and protons.
  The points are 1999 data.
  Right: distribution of \dedx pull for proton hypothesis for electron 
  candidates. 
  The hatched histogram is the background from 
  ${\mathrm e^+e^-\rightarrow e^+e^-q\bar{q}}$ events 
  and the shaded histogram is the rest of the background. 
  The cuts against protons are indicated by arrows. 
  All other selection cuts are applied. The points are 1999 data.} 
\label{dedx}
\end{center}
\end{figure}

A considerable amount of kaon and proton background from 
${ \mathrm e^+e^- \rightarrow e^+e^- q\bar{q}}$ events 
remained after the cuts on the pulls for the muon and electron hypotheses. 
Figure \ref{dedx} (left) shows the specific energy loss for electron candidates
plotted versus the momentum of the particle. 
Proton and kaon bands are clearly visible.
To remove the kaon and proton background, the electron selection was tightened.
The \dedx 
for the electron candidate had to not exceed 1.9 times the minimum ionisation,
and the pulls for the proton and kaon hypotheses for the electron candidate 
both had to be outside the $\pm 1.5 \sigma$ interval: 
$|\Pi_{\mathrm K}|>1.5$ and $|\Pi_{\mathrm p}|>1.5$.
Figure \ref{dedx} (right) shows the distribution of the pull for the proton hypothesis with all 
selection cuts applied except the cut on the variable shown. The hatched histogram shows the background 
from $ {\mathrm e^+e^-\rightarrow e^+e^-q\bar{q}} $ events, 
the shaded histogram shows the rest of the background. The
cuts on this variable are indicated by arrows.

Table~\ref{eff_table} summarises the efficiency of the preselection, 
final step of selection and
overall selection efficiency. The drop in the preselection in 2000
is caused by the removal of events in or near the unstable TPC sector.
The uncertainties of the selection efficiency determination are discussed later 
in the paper.

\begin{table}[h]
  \begin{center}
    \begin{tabular}{|l|c|c|c|c|} 
      \hline 
                 &  1997   &  1998   &  1999  &  2000 \\
      \hline    
    preselection &  5.39   &  5.37   &  5.38  &  3.85 \\     
    selection    &  17.3   &  16.4   &  16.4  &  16.1 \\
      \hline
    overall      &  0.93   &  0.88   &  0.88  &  0.62 \\
      \hline
    \end{tabular}
  \end{center}
  \caption{Efficiencies (\%): preselection, final step of selection and overall efficiency.}
  \label{eff_table}
\end{table}

\begin{figure}[h]
\begin{center}
\begin{tabular}{c c}
 \mbox{\epsfig{file=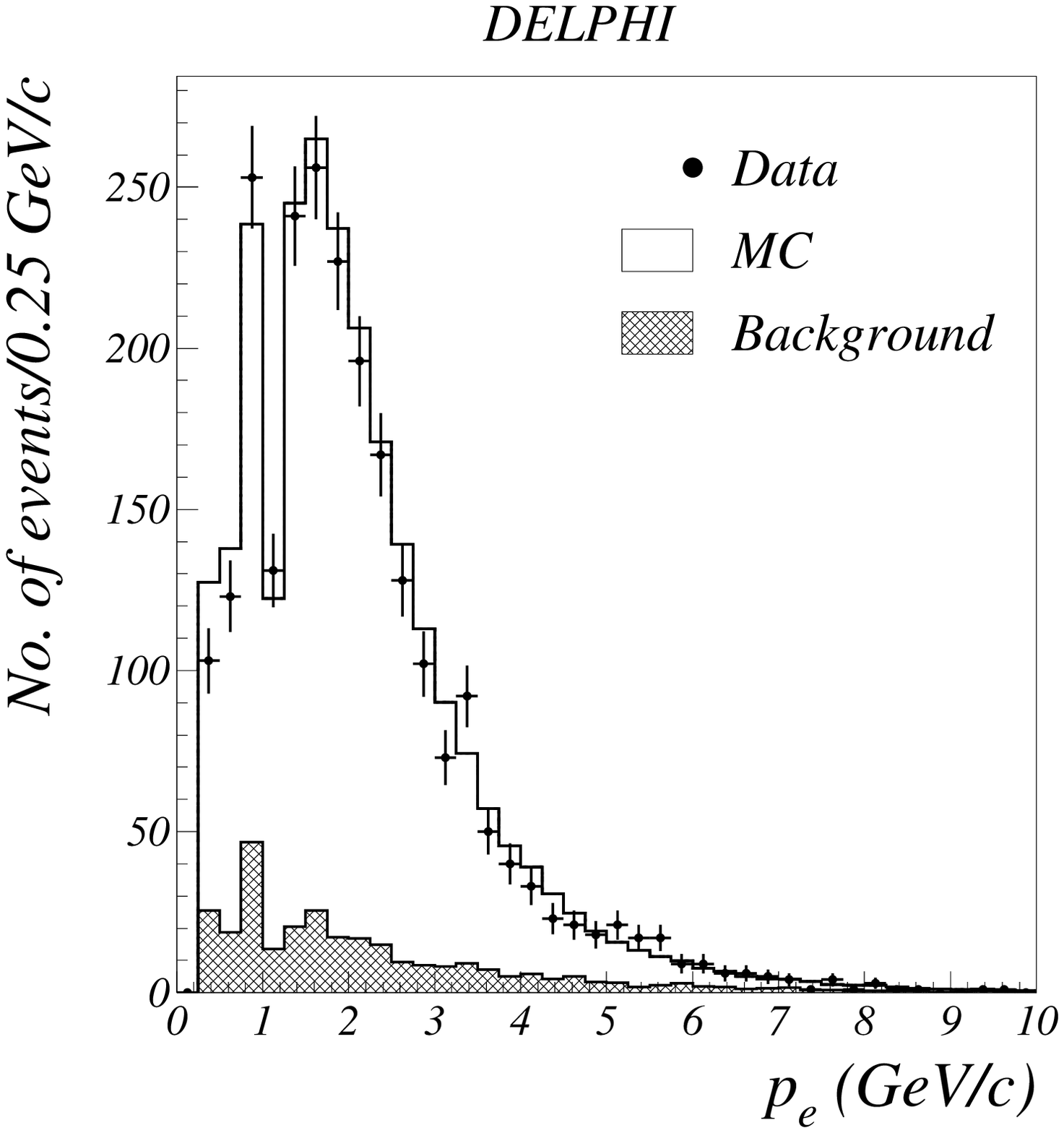,width=0.48\textwidth}} 
 \mbox{\epsfig{file=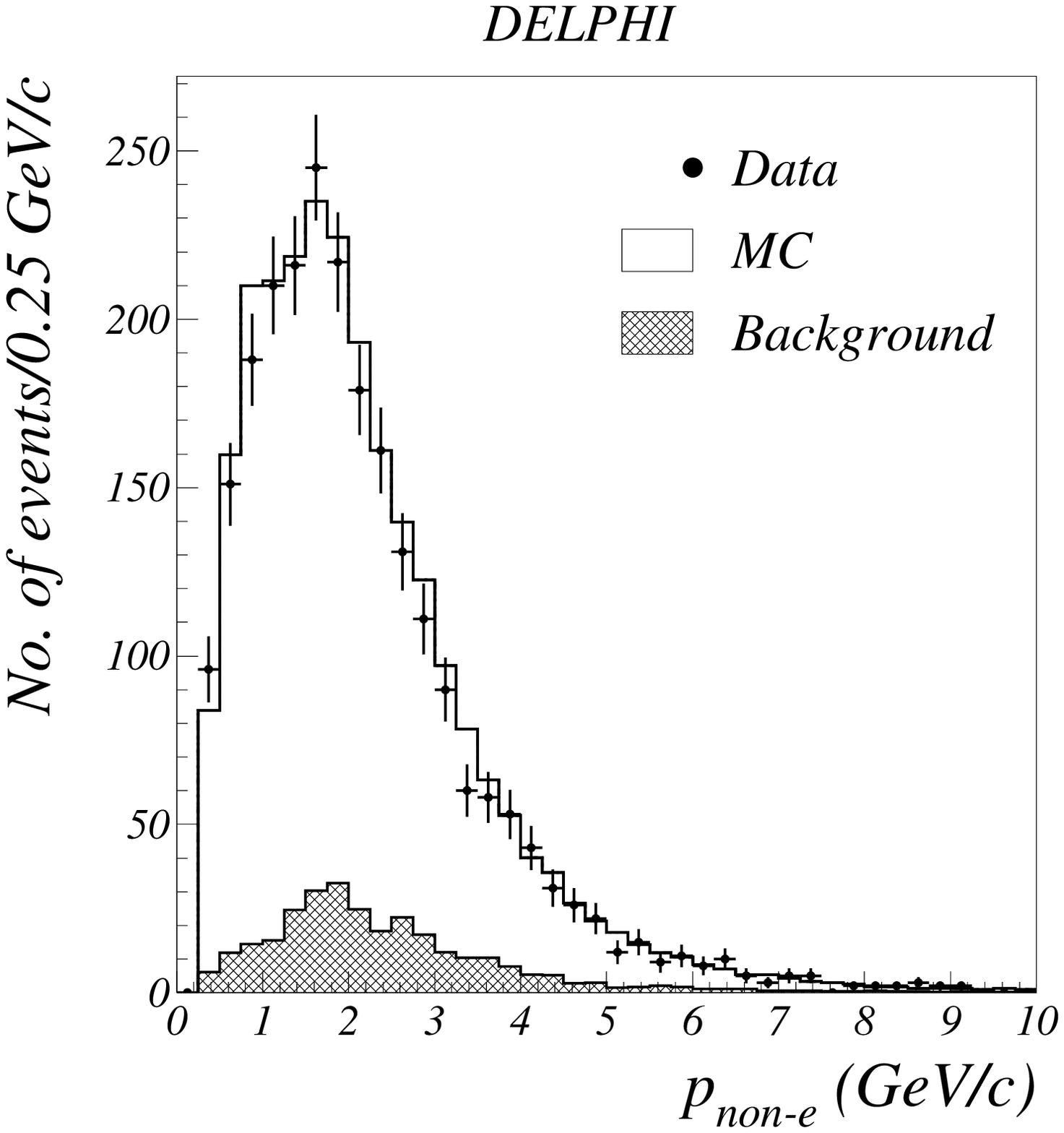,width=0.48\textwidth}} 
\end{tabular}
\caption{Momentum distribution for electron candidates (left) and non-electron candidates (right) 
  for selected events from 1997-2000 data. 
  The distributions of simulated events are corrected for trigger efficiency.  
The dip in the electron momentum distribution near 1 GeV/c is caused by 
the cut against protons: the electron and proton \dedx are equal 
in this region of momentum.}
\label{momenta}
\end{center}
\end{figure}

\begin{figure}[h]
\begin{center}
  \mbox{\epsfig{file=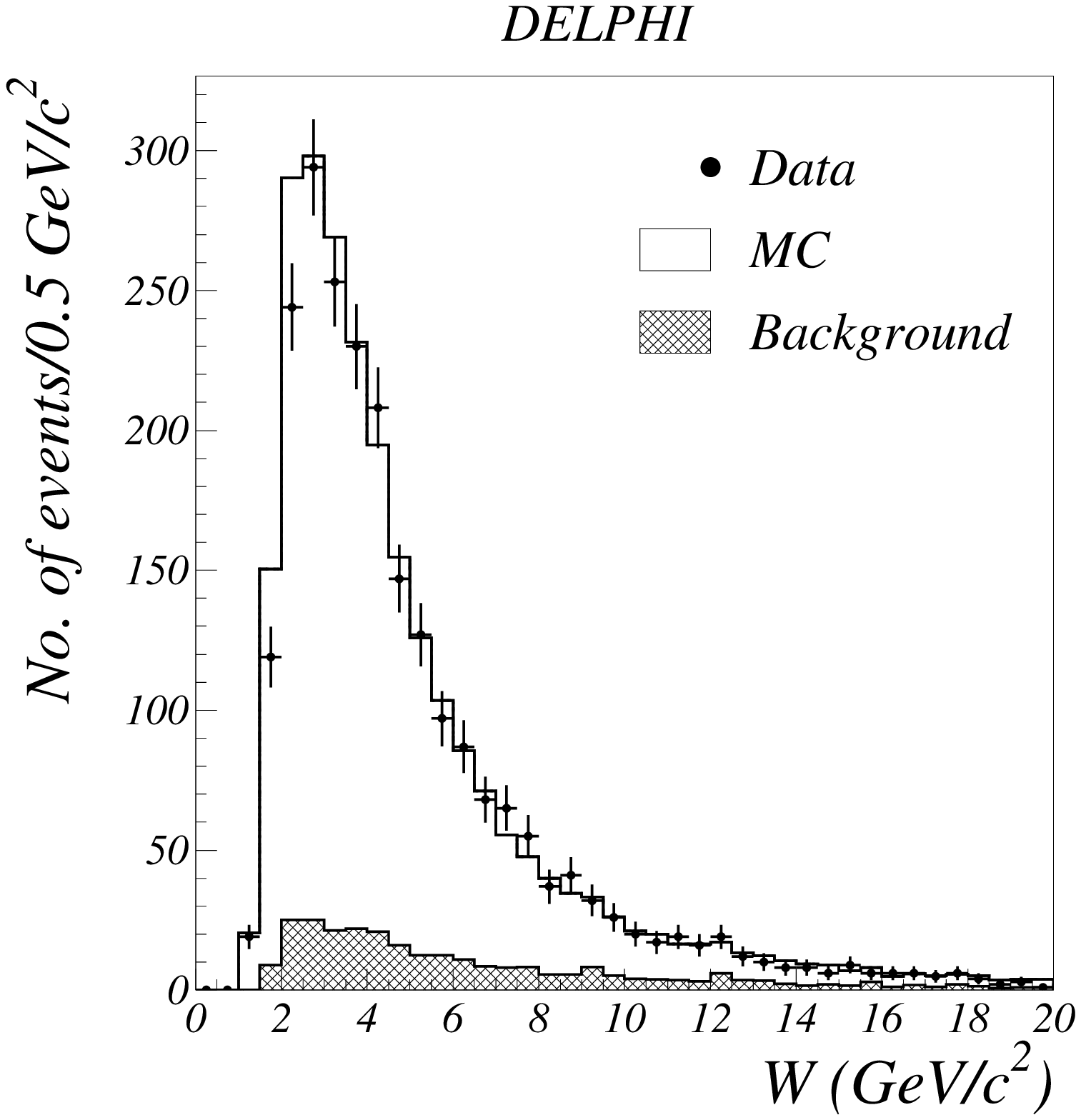,width=0.48\textwidth}} 
\caption{Visible invariant mass distribution for selected events for combined 1997-2000 data.
The distribution of simulated events is corrected for trigger efficiency. The mass was calculated 
using all detected charged particles and photons.
The simulation was corrected for the trigger efficiency.
}
\label{inv.mass}
\end{center}
\end{figure}

In total 2390 candidate events were selected.
Figure \ref{momenta}  compares the distributions of electron and 
non-electron candidate momenta for selected events to the simulation prediction 
for the combined 1997-2000  data. 
Figure \ref{inv.mass} shows the visible invariant mass distribution for selected 
events for the same data sample. Trigger efficiency is taken into account in 
these distributions (see below).


\subsection{Trigger efficiency}
\label{Trigger efficiency}
The low 
momenta of the $\tau$ decay products in the process \signal\  
and the presence of only two tracks in the event could make the probability of triggering on such an
event considerably below 100\%. The determination of the trigger efficiency is therefore
important in this analysis. 

The trigger efficiency was estimated from the subsamples of selected events 
using the fact that an event
can be detected by different components of the DELPHI trigger system \cite{trig}.
Trigger subcomponents of the tracking system were combined into barrel and end-cap triggers. 
For events with one track in the barrel and the other in the end-cap, 
the number of events detected by the barrel trigger ($N_B$), 
by the end-cap trigger ($N_E$), 
and by both triggers ($N_{BE}$) 
were counted using the decision functions of the 
trigger. The barrel and end-cap single track trigger efficiencies were 
calculated, for electrons and non-electrons separately, by the formulae:
\begin{equation}
  \varepsilon_{Barrel} = \frac{N_{BE}}{N_E};  \ \ \ \
  \varepsilon_{end-cap} = \frac{N_{BE}}{N_B}. 
\end{equation}
Finally, the efficiency of the DELPHI calorimetric trigger to the whole event was estimated
from the events triggered by any of the tracking detectors using a similar technique.
The results of the trigger efficiency calculation are summarised in Table~\ref{tref}. 
The track pair trigger efficiency was calculated from the ``OR'' of the single track efficiencies 
using the ratio of the barrel and forward tracks predicted by simulation.
The tau pair trigger efficiency was calculated as ``OR'' of the tracking and calorimetric
triggers.
 
\begin{table}[h]
  \begin{center}
    \begin{tabular}{|ll|c|c|c|c|} 
      \hline 
                      &              & 1997          & 1998        &     1999   &    2000     \\
      \hline
  Barrel track        &              &               &             &            &             \\
                      & electron     &$71.4\pm17.1$  &$94.4\pm5.4$ &$84.6\pm7.1$&$92.3\pm7.4$ \\
                     &non-electron&$100^{+0}_{-17.9}$&$85.2\pm6.8$ &$85.0\pm8.0$&$78.6\pm11.0$\\
    End-cap track     &              &               &             &            &             \\
                      & electron     &$26.3\pm10.1$  &$36.5\pm6.1$ &$21.5\pm4.6$&$22.0\pm6.9$ \\
                      & non-electron &$31.3\pm11.6$  &$30.4\pm6.1$ &$25.9\pm4.8$&$23.1\pm5.8$ \\
      \hline
      Track pair      &         &$94.5^{+5.5}_{-7.1}$&$95.5\pm2.7$ &$93.3\pm3.5$&$93.5\pm4.1$ \\
      Calorimetry     &              &$6.7\pm1.9$    &$8.6\pm1.2$  &$7.1\pm0.9$ &$7.7\pm1.1$  \\
      \hline
      Tau pair        &         &$94.9^{+5.1}_{-6.6}$&$95.9\pm2.5$ &$93.8\pm3.3$&$94.0\pm3.8$ \\
      \hline
    \end{tabular}
    \caption{Summary of the trigger efficiency measurements (\%)}
    \label{tref}
  \end{center}
\end{table}

\subsection{Efficiency of the \dedx measurement}
\label{Selection efficiency correction}
Both tracks of the selected event had to have specific energy loss measurements. An imperfect
detector simulation can lead to a discrepancy in the \dedx measurement efficiency for  good 
tracks in real and simulated events. To take into account this possible disagreement, 
the efficiency for a good track to have a \dedx measurement (to be a ``good TPC track'') 
was calculated for 
${ \mathrm e^+e^- \rightarrow e^+e^-e^+e^-}$ and 
${ \mathrm e^+e^- \rightarrow e^+e^- }\mu^+\mu^-$ samples extracted from preselected events 
(efficiencies to measure \dedx
of pions and muons were assumed to be equal). 
Muon events were selected by requiring at least one track to be 
identified by the muon chambers and electron events were selected using 
information from the DELPHI RICH detectors.
For muon and electron events the efficiency to be a  ``good TPC track'' 
was determined from the ratio
\begin{equation}
\varepsilon^2_{{\mathrm d}E/{\mathrm d}x} = 
\frac{N_{{\mathrm d}E/{\mathrm d}x}}{N_{tot}}
\end{equation}
where $N_{{\mathrm d}E/{\mathrm d}x}$ 
is the number of events with both tracks having a \dedx measurement and 
$N_{tot}$ was the total number of selected events in the given sample.
The tau-pair efficiency was estimated as the product of the single track 
efficiencies for muon and electron.
The tau-pair efficiencies derived from 
${ \mathrm e^+e^- \rightarrow e^+e^- }\mu^+\mu^-$
and ${ \mathrm e^+e^- \rightarrow e^+e^-e^+e^-}$
events for data and Monte Carlo
are presented in Table~\ref{eff}
and were used for selection efficiency correction and for systematic error estimation.
The selection efficiency was multiplied by the factor 
$\frac{\varepsilon_
{{\mathrm d}E/{\mathrm d}x}(data)}
{\varepsilon_{{\mathrm d}E/{\mathrm d}x}(MC)}$ and half of the correction was included
into the systematic error together with the uncertainties from the test sample statistics.
\begin{table}[h]
  \begin{center}
    \begin{tabular}{|l|c|c|c|c|} 
      \hline 
                              & 1997      & 1998        &     1999   &    2000    \\
      \hline
      Efficiency in data, \%  &$82.9\pm1.0$&$82.6\pm0.6$&$82.4\pm0.5$&$83.5\pm0.6$\\
      Efficiency in MC, \%    &$82.3\pm0.4$&$82.5\pm0.2$&$82.3\pm0.1$&$84.6\pm0.2$\\
      \hline
    \end{tabular}
    \caption{Summary of ``good TPC track'' efficiency estimations. These efficiencies are already included 
      in the total efficiency in Table~\ref{eff_table}.}
    \label{eff}
  \end{center}
\end{table}

\subsection{Residual background}
Several sources of background for \signal\  events have been considered:
\begin{itemize}
  \item{The background from 
        ${\mathrm e^+e^- \rightarrow e^+e^-q\bar{q}}$, mainly from protons and kaons
      selected due to the tails of the \dedx pulls for the proton and kaon hypotheses;}
  \item{The background from ${\mathrm e^+e^- \rightarrow e^+e^-e^+e^-}$ 
        and ${\mathrm e^+e^- \rightarrow e^+e^-}\mu^+\mu^-$ events due to the tails of 
	the distributions of the \dedx pulls for the electron and muon hypotheses;}
  \item{
    Background due to other four-fermion processes: 
    non-multiperipheral diagrams (including ${\mathrm e^+e^-}\tau^+\tau^-$ final states)
    and multiperipheral process \signal\ which does not satisfy signal definition;}
  \item{The process ${\mathrm e^+e^-} \rightarrow \tau^+\tau^-$
      (background from other fermion pair production processes was found to be 
      negligible).}
\end{itemize}

The background fractions for the main background sources and their uncertainties
are summarised in Table~\ref{bg}.
The contribution from other background sources was negligible. 
The theoretical precision of ${\mathrm e^+e^- \rightarrow e^+e^-q\bar{q}}$ generation by
PYTHIA is not well known. Therefore it was estimated from the real data by inverting the
\dedx cut on the electron candidate: \dedx$>$1.9 M.I.P. instead of \dedx$<$1.9 M.I.P.
After comparing these test samples enriched
with ${\mathrm e^+e^- \rightarrow e^+e^-q\bar{q}}$ events with the simulation, 
an error of 20\% was ascribed to the ${\mathrm e^+e^- \rightarrow e^+e^-q\bar{q}}$ event generator.

\begin{table}[h]
  \begin{center}
    \begin{tabular}{|l|c|c|c|c|}
      \hline
      Channel                    &  1997      &  1998      &  1999       & 2000          \\
      \hline
      ${\mathrm e^+e^-\rightarrow e^+e^- q\bar{q}}$
      &$4.3\pm1.5$ &$3.2\pm0.7$ &$3.3\pm0.8$  &$3.2\pm0.8$  \\
      ${\mathrm e^+e^-\rightarrow e^+e^-e^+e^-}$      
      &$2.7\pm0.2$ &$3.4\pm0.1$ &$4.0\pm0.1$  &$2.4\pm0.1$  \\
      ${\mathrm e^+e^- \rightarrow e^+e^-}\mu\mu$  
      &$2.9\pm0.1$ &$5.0\pm0.1$ &$2.4\pm0.1$  &$3.8\pm0.1$  \\
      other 4-fermion           
      &$1.5\pm0.3$ &$1.5\pm0.3$ &$1.2\pm0.2$  &$1.2\pm0.2$  \\
      ${\mathrm e^+e^-} \rightarrow \tau\tau$  
      &$0.69\pm0.01$&$0.55\pm0.01$&$0.47\pm0.01$&$0.40\pm0.01$\\
      \hline 
      Total                      
      &$12.1\pm1.5$&$13.6\pm0.8$&$11.4\pm0.8$  &$11.0\pm0.8$  \\
      \hline
    \end{tabular}
    \caption{Summary of background fractions. The numbers are the expected fractions (\%) 
      of the specified backgrounds in the selected sample. Errors
      are statistical errors of the simulated samples and theoretical uncertainties
      of the Monte Carlo generators added in quadrature.}
    \label{bg}
  \end{center}
\end{table}

\section{Systematic error estimation}
The following sources of systematic error on the cross-section measurement
were considered: uncertainties of selection and
trigger efficiencies and uncertainty of background level. 
Track selection, event selection and the statistical error of the simulated samples were 
taken into account in the calculation of the uncertainty in the selection efficiency.  

The systematic error arising from track selection was estimated in the following way. 
Each cut of the track selection was varied typically by the size of the resolution of the corresponding 
variable from its nominal value in both directions.
The corresponding change of the cross-section $\Delta$ was compared to the value of 
the expected statistical fluctuation $\sigma$ due to the non-identical event sample. 
If the value $\Delta$ was less than $\sigma$,
no systematic error was ascribed to the corresponding cut; in the opposite case
the value of  $\sqrt{\Delta^2-\sigma^2}$ was included in the systematic error. 
The systematic error arising from varying the event selection cuts was estimated in a similar way.

To calculate the systematic error due to the angular corrections applied to the
\dedx measurements, the
\dedx correction functions were varied by the uncertainty of each of their parameters and the analysis 
chain was repeated. The variation of the measured cross-section was added to the systematic error. 
The systematic errors corresponding to scaling and smearing the pulls were calculated similarly. 

The systematic errors associated with track selection cuts, event selection cuts 
and \dedx corrections are summarised in Table~\ref{systematics1}. 
The numbers are given for 1999 data (for other years 
uncertainties of most of the sources scale approximately as inverse square root 
of the statistics).
Additional contributions to the 
selection efficiency uncertainty
also presented in Table~\ref{systematics1} 
come from the statistical error of the Monte Carlo
sample and the selection efficiency correction described in Section 
\ref{Selection efficiency correction}. 

\begin{table}[ht]
  \begin{center}
    \begin{tabular}{|l|l|c|} 
      \hline 
      syst. error source    &                     &         value, \% \\
      \hline
      track selection cuts  &                     &                   \\
                        & $R_{imp}$           &           0.7     \\
                        & $Z_{imp}$           &           1.1     \\
                        & $\delta p/p$        &           0.7     \\
      event selection  cuts &                     &                   \\
                        & $\Pi_{\mathrm e}$           &           0.3     \\
                        & $\Pi_{\mu}$         &           0.3     \\
                        & acoplanarity       &           0.6     \\
\dedx corrections &                     &                   \\
                        &$\Pi_{\mathrm e}$ $\theta$   &           1.0     \\
                        &$\Pi_{\mathrm e}$ $\phi$     &           0.9     \\
                        &$\Pi_{\mu}$ $\theta$ &           1.0     \\
                        &$\Pi_{\mu}$ $\phi$   &           1.0     \\
                        & scaling             &           0.7     \\
                        & smearing            &           0.6     \\
      MC statistics     &                     &           0.8     \\
      ``Good TPC track''&                     &                   \\
      ~~~correction     &                     &           0.6     \\
    \hline
    Total               &                     &           3.0     \\
    \hline
    \end{tabular}
    \caption{Systematic errors for 1999 data coming from track selection, 
      event selection, \dedx corrections, simulated samples statistics and
      ``good TPC track'' correction.}
    \label{systematics1}
  \end{center}
\end{table}   

The largest contribution to the systematic error is given by the uncertainty of the 
trigger efficiency determination, 
dominated by the statistics of the real data events, 
see Section \ref{Trigger efficiency} and Table~\ref{tref}.
An additional contribution arises because the trigger efficiency 
for background events, assumed to be equal to that of the signal, may be different.
A conservative estimate of this uncertainty was obtained by changing the trigger efficiency
for background upwards to 100\% and downwards by the same amount.

The systematic error due to residual background includes
the simulated sample statistical uncertainty and the theoretical uncertainty of 
the Monte Carlo generators, mainly
for the ${\mathrm e^+e^- \rightarrow e^+e^-q\bar{q}}$ process, see Table~\ref{bg}.

The sources of selection efficiency uncertainty are described in detail in
Table~\ref{systematics1}.

The sources of systematic uncertainty
are summarised in Table~\ref{systematics2}. 
Total systematic errors 
calculated as the sum in quadrature of all described components are also 
presented in Table~\ref{systematics2}. The following uncertainties
were assumed to be fully correlated between different years: generator theoretical error; 
trigger efficiency for background; and uncertainties estimated from 
variation of track and event selection cuts. Systematic errors from 
other sources were treated as uncorrelated.

\begin{table}[h]
  \begin{center}
    \begin{tabular}{|l|c|c|c|c|} 
      \hline 
                       &1997     & 1998        &     1999   &    2000   \\
      \hline    
      Trigger eff.     & 7.0     & 2.7         &     3.6    &    4.5    \\
      Selection eff.   & 5.1     & 3.2         &     3.0    &    3.0    \\
      Background       & 1.7     & 0.9         &     0.9    &    0.9    \\
      Luminosity       & 0.6     & 0.6         &     0.6    &    0.6    \\
      \hline
      Total            & 8.9     & 4.3         &     4.7    &    5.4    \\
      \hline
    \end{tabular}
    \caption{Relative systematic errors 
      on cross-section (in \%).}
    \label{systematics2}
  \end{center}
\end{table}

\section{Results of the cross-section measurement}
%

The cross-sections were computed using the formula
\begin{equation}
  \sigma = \frac{N_{obs}-N_{bg}}{\varepsilon_{sel}\varepsilon_{trig}{\cal L}}
\end{equation}
where $N_{obs}$ is the number of observed events, $N_{bg}$ is the expected number of 
background events in the assumption that background events have the same 
trigger efficiency as signal events, 
$\varepsilon_{sel}$ is the selection efficiency, $\varepsilon_{trig}$
is the trigger efficiency and $\cal L$ is the integrated luminosity.

The numbers of observed and expected events, 
the measured cross-sections and the cross-sections from the BDKRC Monte Carlo
simulation together with their ratios are presented in Table~\ref{cross-section}. 
The predicted number of events was calculated from
the signal and background simulation, taking into account
trigger efficiency and corrections to the \dedx efficiency.
Agreement was found between the measurements and the  Standard Model (SM) 
predictions calculated by BDKRC.
The ratio of observed and predicted cross-sections was averaged over all LEP2 data,
taking into account correlations of systematic errors. The result
was found to be $0.96\pm0.04$. The average LEP2 cross-section
is $429\pm 17$ pb 
corresponding to the luminosity-weighted mean centre-of-mass energy of 
$197.1$ GeV.
The cross-section predicted at this energy by BDKRC is $447.7\pm0.3$ pb.

\begin{table}[h]
  \begin{center}
    \begin{tabular}{|c|c|c|c|c|c|} 
      \hline 
       Year & Observed  & Expected   &$\sigma_{meas}$, pb& $\sigma_{MC}$, pb &$\sigma_{meas}/\sigma_{MC}$  \\
      \hline                                                                                    
      1997  &$  211    $&$ 224\pm18 $&$ 401\pm32\pm 36 $&$428.2\pm0.5       $ &$0.94\pm0.11$\\
      1998  &$  629    $&$ 652\pm24 $&$ 419\pm19\pm 18 $&$436.7\pm0.5       $ &$0.96\pm0.06$\\
      1999  &$  909    $&$ 937\pm39 $&$ 436\pm16\pm 21 $&$448.5\pm0.5       $ &$0.97\pm0.06$\\
      2000  &$  641    $&$ 665\pm32 $&$ 443\pm20\pm 24 $&$459.4\pm0.5       $ &$0.97\pm0.07$\\
      \hline
    \end{tabular}
    \caption{The numbers of observed and expected events, 
      measured cross-sections, QED predictions and their ratios. The first error on the measured
      cross-sections is statistical, the second is systematic.}
    \label{cross-section}
  \end{center}
\end{table}

\section{Determination of anomalous magnetic and electric dipole moments}
In the Standard Model, leptons are considered as point-like objects. 
Therefore the observation of a deviation of the magnetic or electric dipole
moments of the  leptons from their SM values would open a window onto the
physics beyond the SM.
The anomalous magnetic moments of the electron \cite{e} and muon \cite{mu}
are known with high precision, but the short life-time of the tau-lepton 
does not allow measurement of its anomalous moments with similar precision by
a spin precession method.

The generalised form of the $\tau\tau\gamma$ vertex can be parametrised 
as follows:
\begin{equation}
 -ie\bar{u}(p')\{
                 F_1(q^2)\gamma^{\mu} + 
                iF_2(q^2)\sigma^{\mu\nu}\frac{q_{\nu}}{2m_{\tau}} +
                 F_3(q^2)\gamma^5\sigma^{\mu\nu} \frac{q_{\nu}}{2m_{\tau}}\}
u(p)\epsilon_{\mu}(q) 
\label{anomal}
\end{equation}
where $\epsilon_{\mu}(q)$ is the polarization vector of the photon with momentum $q$.
The form factor $F_1$ describes the distribution of electric charge and 
$e_{\tau} = eF_1(0)$, while $F_2$ and $F_3$ are form factors related to the anomalous 
magnetic moment $a_{\tau}$ and electric dipole moment $d_{\tau}$:
\begin{equation}
  a_{\tau} \equiv \frac{g_{\tau}-2}{2}=F_2(0)
\end{equation}
and 
\begin{equation}
  F_3(0)=-\frac{{2m_{\tau}} d_{\tau}}{e_{\tau}}
\end{equation}
In the SM at tree level, $a_{\tau} = 0$ and $d_{\tau} = 0$. Accounting for loop
diagrams gives a non-zero value to $a_{\tau} = 11773(3)\cdot 10^{-7}$ 
\cite{samuel}, while a non-zero value of $d_{\tau}$ is forbidden by both T invariance and P invariance.

The values of $a_{\tau}$ and $d_{\tau}$ have been measured by several groups.
The L3 and OPAL collaborations \cite{l3,opal} studied radiative 
$Z\rightarrow \tau\tau\gamma$ events and set the following 95 \% CL limits 
on the values of the anomalous magnetic and electric dipole moments:
\begin{center}
  \begin{tabular}{lll}
    $-0.052<a_{\tau}<0.058$ & and $|d_{\tau}|<3.1~~ ( 
                              10^{-16}$ $e\cdot \mathrm{cm})$ & (L3) \\
    $-0.068<a_{\tau}<0.065$ & and $|d_{\tau}|<3.7~~ (
                              10^{-16}$ $e\cdot \mathrm{cm})$ & (OPAL).
  \end{tabular}
\end{center}

The best limit so far on $d_{\tau}$ was obtained by BELLE \cite{belle}:

$$-0.22 < \Re e(d_{\tau}) < 0.45 ~~ (10^{-16}e\cdot \mathrm{cm})$$
$$-0.25 < \Im m(d_{\tau}) < 0.08 ~~ (10^{-16}e\cdot \mathrm{cm})$$

Other limits on $a_{\tau}$ and $d_{\tau}$ can be found in \cite{gonzalez}.  

%
%


\subsection{Limits from this analysis}
Here we present the study of the anomalous magnetic and electric dipole 
moments of the tau lepton based on the analysis of the \signal\  cross-section.
The study of anomalous couplings of tau leptons to photons at LEP 
in this channel was proposed in \cite{cornet_paper}.

To model the contribution of non-SM anomalous magnetic and dipole moments 
we use the calculation by Cornet and Illana \cite{cornet_privat}. The calculation is based
on computation of the matrix element of the process $\gamma\gamma \rightarrow
\tau^+\tau^-$ in leading order of QED and its translation to the cross-section
of the \signal\  process using the Equivalent Photon Approximation (EPA) \cite{budnev}.
The EPA parameter $(-q^2)_{max}$ (the upper limit of the 
integration over 4-momenta of the emitted photon)  was chosen such that the total 
cross-section predicted by EPA (with SM values of anomalous 
electromagnetic moments) agreed with BDKRC calculation.
According to the calculations \cite{cornet_privat} each of the anomalous
terms of (\ref{anomal}) would mainly modify the rate of tau pair                    
production in the barrel region of the detector                                   
where the experimental selection has largest efficiency.                        
This leads to a larger selection efficiency for                                 
the anomalous term contribution, improving in principle the                           
limits obtained on anomalous moments. However in this paper we conservatively                           
assumed that the standard and anomalous contributions                           
have the same selection efficiency.

\begin{figure}[h]
\begin{center}
\begin{tabular}{c c}
 \mbox{\epsfig{file=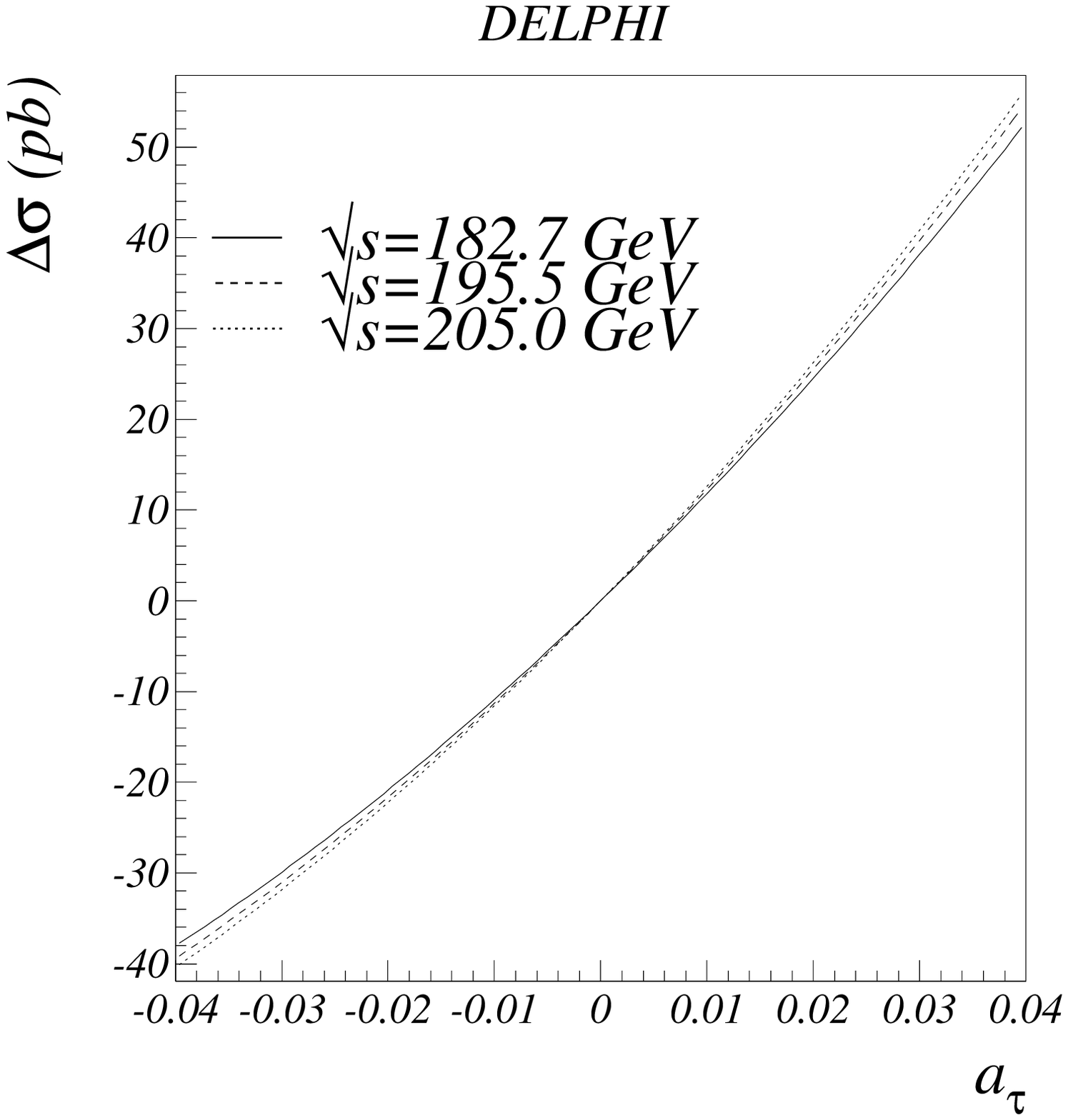,width=0.48\textwidth}} 
 \mbox{\epsfig{file=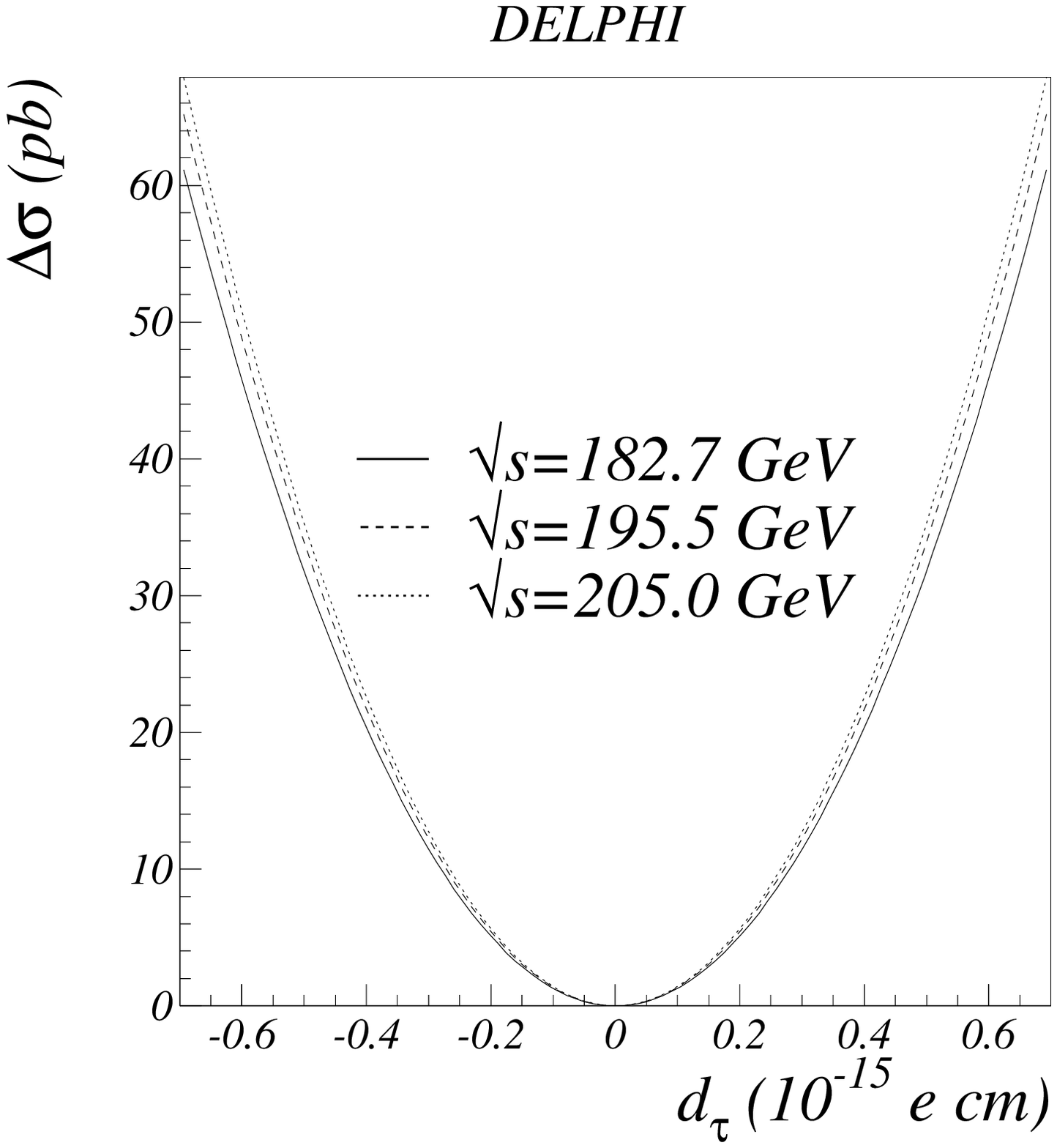,width=0.48\textwidth}} 
\end{tabular}
\caption{Total cross-section change as a function of anomalous
magnetic moment and as a function of electric dipole moment.}
\label{xs}
\end{center}
\end{figure}

Figure \ref{xs} shows how the total cross-section changes as a function of the
anomalous magnetic moment and as a function of the electric dipole moment. The
three lines on each plot represent the calculation with $\sqrt{s}=$182.7, 195.5
and 205.0 GeV. Increasing the collision energy slowly increases both non-SM
contributions. However, increasing the magnitude of the anomalous magnetic
moment can either increase or decrease the  cross-section while increasing that
of the electric dipole moment tends only to increase the  cross-section.

To compare the experimentally measured values of the cross-sections 
to the non-SM calculation, they were first converted
from accepted to the total cross-sections, taking into account the 1.44\%
difference due to the signal definition (see section 2). 
The validity of applying SM conversion factors
is supported by the fact that the measured cross-sections are in good agreement 
with the SM prediction, which guarantees the smallness of the non-SM contribution,
and by the fact that the correction itself is small. 

Fits to the cross-sections measured in 1997, 1998, 1999 and 2000 were performed 
taking  $a_{\tau}$ and $d_{\tau}$ as parameters. When 
fitting for $a_{\tau}$, the value of $d_{\tau}$ was set to its SM value and 
{\it vice versa}. The errors on the cross section measurements were taken as 
the statistical and systematic errors added in quadrature. 

To quote the obtained limits we used the following 
convention:
\begin{equation}
  \int_{-\infty}^{L}\exp{(-\chi^2/2)}\,da_{\tau} =   
   \int_{R}^{ \infty}\exp{(-\chi^2/2)}\,da_{\tau} = \frac{1-CL}{2} 
\end{equation}
where $CL$ is the desired confidence level and $L$ and $R$ 
are lower and upper limits. A similar definition was used for $d_{\tau}$.
We quote central values $\mu$ and errors $\sigma$ 
for moments according to
\begin{equation} 
  \sigma = \frac{R-L}{2},\ \ \  \mu = \frac{R+L}{2}.
\end{equation}
where $R$ and $L$ are calculated with \mbox{68.3\%}  confidence level.

\begin{figure}[h]
\begin{center}
\begin{tabular}{c c}
 \mbox{\epsfig{file=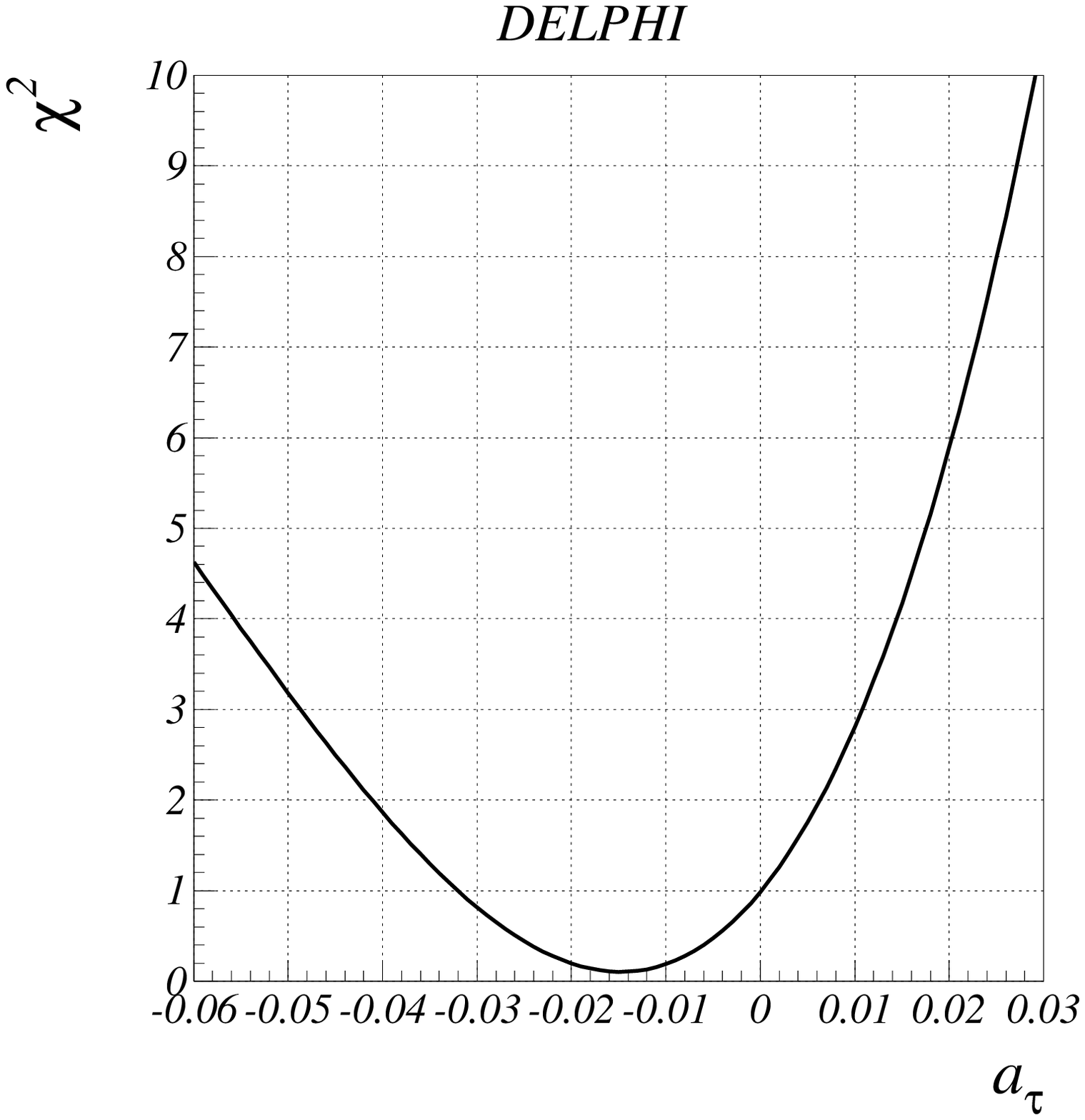,width=0.48\textwidth}} 
 \mbox{\epsfig{file=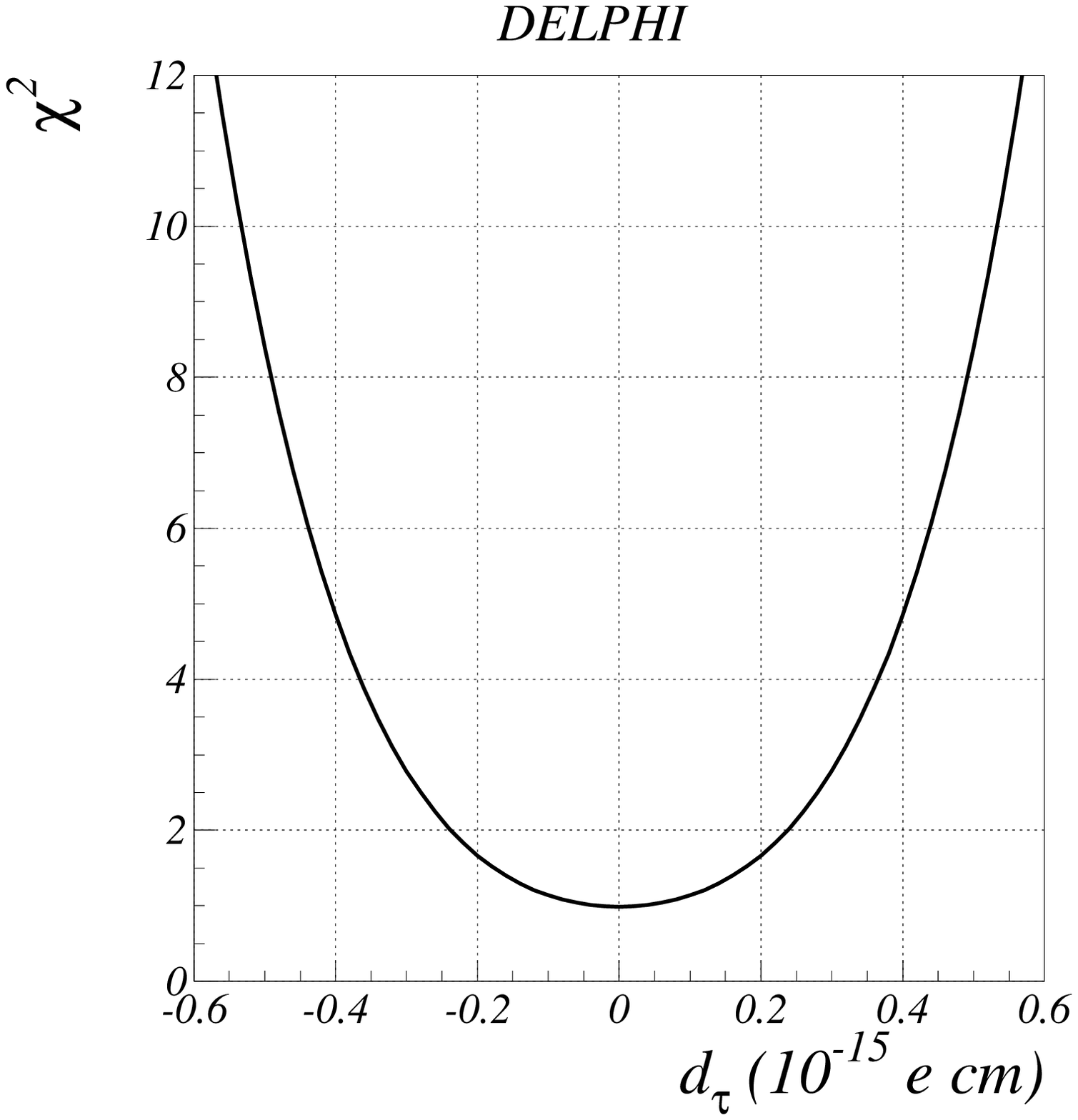,width=0.48\textwidth}} 
\end{tabular}

\caption{ $\chi^2$ as a function of anomalous magnetic moment and 
as a function of electric dipole moment. }
\label{chi2}
\end{center}
\end{figure}

Figure \ref{chi2} shows the $\chi^2$ as a function of the anomalous magnetic moment and 
as a function of the electric dipole moment. The results of the fit are:
\begin{center}
  \begin{tabular}{ll}
    $-0.052<a_{\tau}<0.013$, & 95\% CL\\
    $|d_{\tau}|<3.7\cdot 10^{-16}$ $e\cdot \mathrm{cm}$, & 95\% CL.
  \end{tabular}
\end{center}

\noindent The limit on $a_{\tau}$ improves the current PDG limit \cite{PDG}
based on the L3 result \cite{l3}.

\begin{figure}[h]
\begin{center}
  \mbox{\epsfig{file=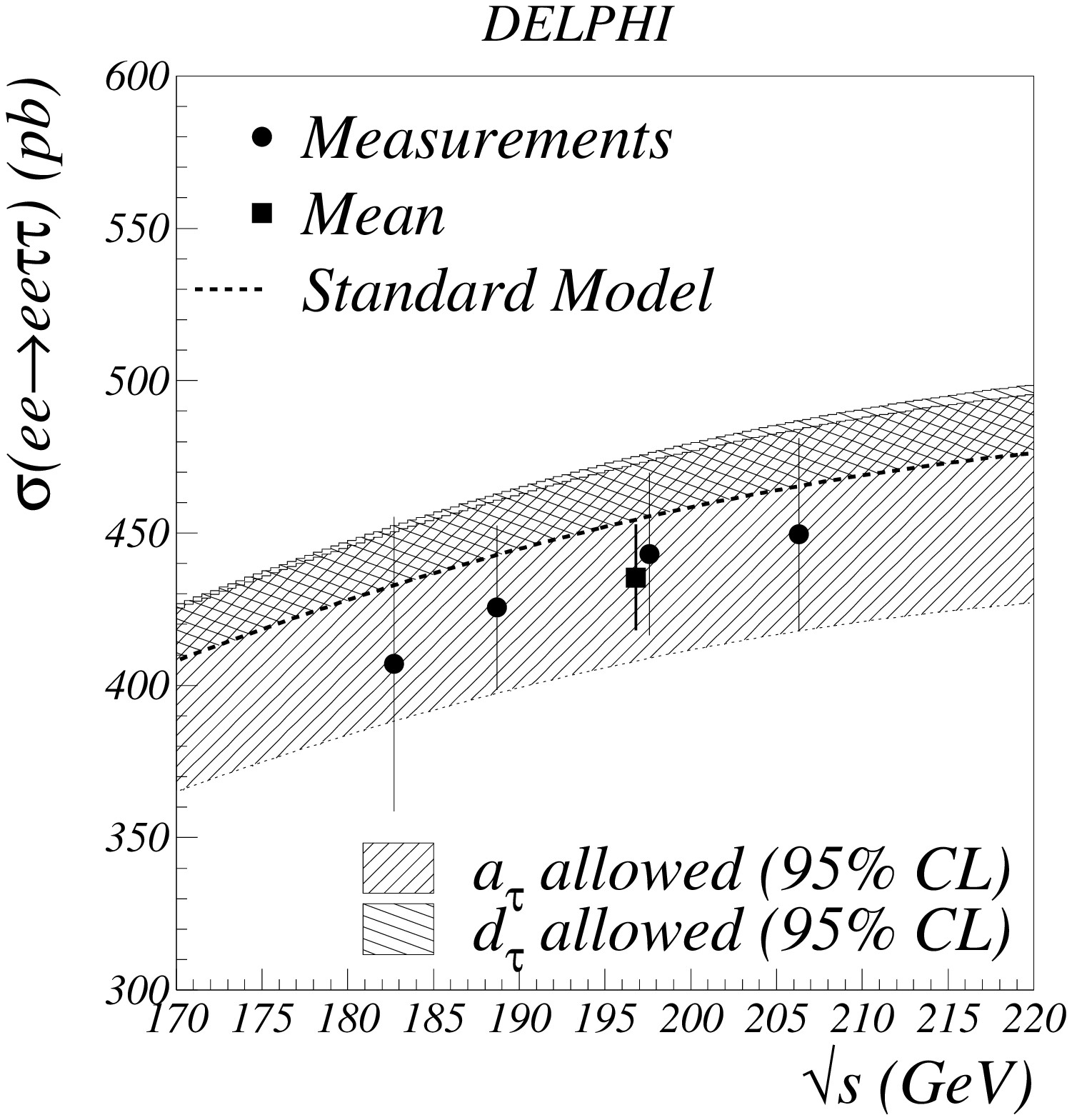,width=0.60\textwidth}} 
\caption[]{Measured cross-section (circles), average LEP2 cross-section (square) and 
  SM expectation as a function of $\sqrt{s}$.  The two bands show the cross-section 
  variation allowed due to anomalous magnetic
  and electric dipole moments within 95\% limits from this analysis. }
  \label{points}
\end{center}
\end{figure}

Figure \ref{points} shows the the measured cross-section, average LEP2 cross-section and 
SM expectation as a function of $\sqrt{s}$. Two bands superimposed on the plot
represent the allowed region for the cross-section variation due to anomalous magnetic
and electric dipole moments. 
The results 
expressed in the form of central value and error are the following:

\begin{center}
$a_{\tau}=-0.018\pm0.017$, \\
$d_{\tau}=(0.0\pm2.0) \cdot 10^{-16}$ $e\cdot \mathrm{cm}$.
\end{center} 

\section{Conclusion}
We have studied the reaction \signal\ with the data 
collected with the DELPHI detector during LEP2 operation in the years 1997-2000.
The average LEP2 cross-section was found to be 429$\pm$17 pb compared
to 447.7 pb expected from the Standard model. The measured/predicted ratio
0.96$\pm$0.04 agrees with the QED prediction at the level of one standard deviation.
The measured cross-sections were used to extract limits on the anomalous
magnetic and electric dipole moments of the tau lepton. The 95\% CL limits obtained 
are 
\begin{center}
    $-0.052<a_{\tau}<0.013$ \\
    $|d_{\tau}|<3.7\cdot 10^{-16}$ $e\cdot \mathrm{cm}$.
\end{center} 

\subsection*{Acknowledgements}
\vskip 3 mm
 We are greatly indebted to our technical 
collaborators, to the members of the CERN-SL Division for the excellent 
performance of the LEP collider, and to the funding agencies for their
support in building and operating the DELPHI detector.\\
We acknowledge in particular the support of \\
Austrian Federal Ministry of Education, Science and Culture,
GZ 616.364/2-III/2a/98, \\
FNRS--FWO, Flanders Institute to encourage scientific and technological 
research in the industry (IWT), Federal Office for Scientific, Technical
and Cultural affairs (OSTC), Belgium,  \\
FINEP, CNPq, CAPES, FUJB and FAPERJ, Brazil, \\
Czech Ministry of Industry and Trade, GA CR 202/99/1362,\\
Commission of the European Communities (DG XII), \\
Direction des Sciences de la Mati$\grave{\mbox{\rm e}}$re, CEA, France, \\
Bundesministerium f$\ddot{\mbox{\rm u}}$r Bildung, Wissenschaft, Forschung 
und Technologie, Germany,\\
General Secretariat for Research and Technology, Greece, \\
National Science Foundation (NWO) and Foundation for Research on Matter (FOM),
The Netherlands, \\
Norwegian Research Council,  \\
State Committee for Scientific Research, Poland, SPUB-M/CERN/PO3/DZ296/2000,
SPUB-M/CERN/PO3/DZ297/2000 and 2P03B 104 19 and 2P03B 69 23(2002-2004)\\
JNICT--Junta Nacional de Investiga\c{c}\~{a}o Cient\'{\i}fica 
e Tecnol$\acute{\mbox{\rm o}}$gica, Portugal, \\
Vedecka grantova agentura MS SR, Slovakia, Nr. 95/5195/134, \\
Ministry of Science and Technology of the Republic of Slovenia, \\
CICYT, Spain, AEN99-0950 and AEN99-0761,  \\
The Swedish Natural Science Research Council,      \\
Particle Physics and Astronomy Research Council, UK, \\
Department of Energy, USA, DE-FG02-01ER41155, \\
EEC RTN contract HPRN-CT-00292-2002. \\
We thank F.Cornet for providing the calculation of the cross-section
of anomalous tau-pair production in two-photon collisions.


\end{document}